\RequirePackage{fix-cm}
\documentclass[twocolumn]{svjour3}          
\usepackage{amsmath}
\usepackage{algpseudocode}
\usepackage{tabularx, subfigure}
\usepackage{graphicx}
\usepackage{amssymb}
\usepackage{amsmath,epsfig,amsfonts,amssymb,graphicx}
\usepackage{algpseudocode}
\usepackage{multirow}
\usepackage{cite}
\usepackage{wasysym}
\usepackage{marvosym}
\usepackage{amssymb}
\usepackage{url}
\usepackage[linesnumbered,ruled,vlined]{algorithm2e}
\providecommand{\SetAlgoLined}{\SetLine}

\smartqed  
\usepackage{graphicx}
\newtheorem{thm}{Theorem}
\newtheorem{defn}{Definition}

\begin{document}

\title{HFUL: A Hybrid Framework for User Account Linkage across Location-Aware Social Networks}


\author{Wei Chen $^1$ \and Weiqing Wang$^2$ \and Hongzhi Yin$^3$ \and Lei Zhao$^{1}$(\Letter) \and Xiaofang Zhou$^4$ 
}


\institute{Wei Chen \at robertchen@suda.edu.cn \\ 
	Weiqing Wang \at teresa.wang@monash.edu \\ Hongzhi Yin \at y.yin1@uq.edu.au  \\
	Lei Zhao \at zhaol@suda.edu.cn\\  Xiaofang Zhou \at zxf@cse.ust.hk \\
	$^1$Institute of Artificial Intelligence, School of Computer Science and Technology, Soochow University, China\\
	$^2$Faculty of Information Technology, Monash University, Melbourne, Australia\\
	$^3$School of ITEE, The University of Queensland, Bribane, Australia\\
	$^4$The Hong Kong University of Science and Technology, Hong Kong, China
}

\date{Received: date / Accepted: date}

\maketitle

\begin{abstract}
Sources of complementary information are connected when we link user accounts belonging to the same user across different platforms or devices. The expanded information promotes the development of a wide range of applications, such as cross-platform prediction, cross-platform recommendation, and advertisement. Due to the significance of user account linkage and the widespread popularization of GPS-enabled mobile devices, there are increasing research studies on linking user account with spatio-temporal data across location-aware social networks. Being different from most existing studies in this domain that only focus on the effectiveness, we propose a novel framework entitled HFUL (A \textbf{H}ybrid \textbf{F}ramework for \textbf{U}ser Account \textbf{L}inkage across Location-Aware Social Networks), where efficiency, effectiveness, scalability, robustness, and application of user account linkage are considered. Specifically, to improve the efficiency, we develop a comprehensive index structure from the spatio-temporal perspective, and design novel pruning strategies to reduce the search space. To improve the effectiveness, a kernel density estimation-based method has been proposed to alleviate the data sparsity problem in measuring users' similarities. Additionally, we investigate the application of HFUL in terms of user prediction, time prediction, and location prediction. The extensive experiments conducted on three real-world datasets demonstrate the superiority of HFUL in terms of effectiveness, efficiency, scalability, robustness, and application compared with the state-of-the-art methods.
\keywords{User Account Linkage \and Social Networks \and Location Data}
\end{abstract}

\section{Introduction}\label{1:introduction}

The proliferation of GPS-enabled devices, such as vehicles, mobile phones, and smart bracelets, leads to the increasing availability of location data from two perspectives: 1) the volume of location data increases unprecedentedly; 2) the resources of location data tend to be more diverse. Recently, much more location data has been  generated by newly-emerging location-aware social networks (LBSNs), such as Foursquare, Twitter, and Instagram. Many users have registered accounts on these platforms, and posted their statuses associated with location information, referred as ``check-ins''. Compared with other online activities, such as commenting, tagging, and following, ``check-ins'' bridge the gap between the real world and the virtual world with the geographical data \cite{gao2014data}. The study of check-in data provides an unprecedented opportunity to analyze users' real world behaviors and potentially improve a variety of location-aware services \cite{pham2013ebm}\cite{lichman2014modeling}. For example, in \cite{Riederer2016Lingking}, check-in records are used to link user accounts across different platforms. Obviously, compared with the information collected from one specific platform, we can obtain more comprehensive user information after identifying and linking user accounts across platforms, since the sources of complementary information are integrated. From a commercial perspective, this expanded information will benefit many location-aware applications, such as maps and cross-platform recommendation. Consequently, linking user accounts across location based social networks has attracted increasing attention. However, despite of the significance of the study, following inevitable problems bring great challenges for this work.

\subsection{Challenges}\label{sec:challenge}
\textbf{Data Sparsity}. The density of the check-in data for each user is of critical importance to  user account linkage across LBSNs. This is because we can model a user's real behaviors more precisely with a denser dataset, which enables us to link user accounts across different platforms more accurately. Unfortunately, user-generated  check-in datasets are extremely sparse. Compared with the traditional GPS datasets, where users' geographical locations are automatically recorded by the GPS devices and the time periods between two consecutive points are usually short, the check-in process is user-driven on location-aware social networks, i.e., users decide whether to check in at a specific place or not. Such user-driven mechanism leads to the data sparsity problem from the following  aspects. First, the number of check-in records generated by each user is rather limited, as many users are reluctant to post their statuses \cite{noulas2011empirical}. Second, the spatial span of check-in data is extremely large \cite{gao2014data}. For example, a user may usually check in at Boston, but with the latest check-ins at California. Third, the time spans between consecutive check-ins are usually wide, where some users even have more than one-year gaps between consecutive check-ins \cite{gao2014data}\cite{wang2016spore}. All these behaviors lead to the significant sparsity of geographical data on location-aware social networks, and bring enormous difficulty for precise linkage.

\begin{figure}[h]
	\centering
	\includegraphics[width=0.49\textwidth]{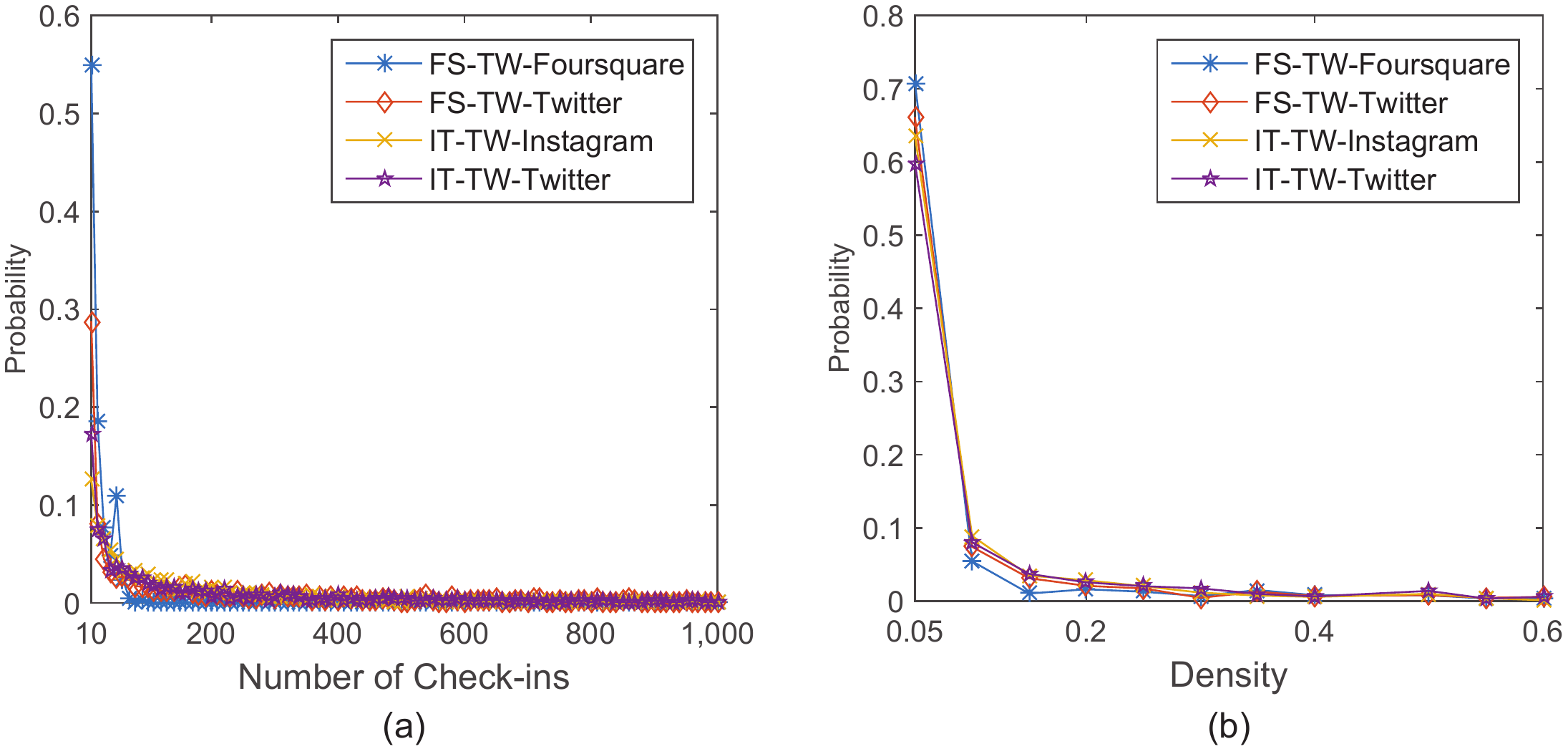}\\
	\caption{(a) The distribution of the number of check-ins; (b) The distribution of density, i.e., the number of check-ins/km$^2$}
	\label{1:data sparsity}
	\vspace{-5pt}
\end{figure}

To illustrate the data sparsity problem of user check-ins more clearly, we conduct an analysis on two real cross-platform datasets, i.e., the dataset Foursquare-Twitter (FS-TW) provided by \cite{zhang2014transferring}\cite{Riederer2016Lingking} and the dataset Instagram-Twitter (IT-TW) provided by \cite{Riederer2016Lingking}. The analysis results are presented in Figure \ref{1:data sparsity}. The distribution of the average number of check-in records is presented in Fig. \ref{1:data sparsity}(a). Obviously, most users of the given datasets have a small volume of check-in records (i.e., usually less than 50 check-in records). Fig. \ref{1:data sparsity}(b) reveals the data sparsity problem from a different perspective, where the density is defined as the number of check-in records per km$^2$. For most users, they have less than 0.2 check-ins in one km$^2$.

\textbf{Data Imbalance.} Data imbalance in user account linkage refers to the phenomenon that, the number of check-in records of the same user is different across different platforms. This is because a user is not
likely to post the check-in about the same activity many times across different platforms, i.e.,  some of a user's check-in records are missing on certain platforms. For example, many users have registered accounts on Facebook, Twitter, and Instagram simultaneously, yet they may just select one platform to post a check-in after taking an activity in a venue or location. Moreover, users' consistent preferences make the situation worse. Users' consistent preferences refer to that users may prefer one platform to post their check-ins consistently, which can be caused by various factors (e.g., the influence of social friends). For example, if most of a user's friends use Facebook, he/she may always prefer posting check-ins on Facebook, which means that most of his/her activities are missing on other platforms.

\textbf{Negative Coincidence.} Negative coincidence occurs when two different user accounts happen to have check-in records at same places many times \cite{pham2013ebm}. Such phenomenon tends to happen at popular and crowded places (e.g., supermarkets, cafeteria, and schools), where many users tend to visit repeatedly \cite{li2010mining}\cite{Yuan2013WhoWW} and share their statuses associated with location data. Although many records are generated at these places, we cannot distinguish users from each other, due to the low discrimination of these check-ins.

\textbf{Low Veracity} The instability of GPS-enabled devices leads to the low quality of spatio-temporal data, where there usually exist some outliers. Obviously, data outlier is a big negative factor for precise user linkage. Unfortunately, such a phenomenon is ubiquitous across different location-aware social networks.

\subsection{Solutions}
A straightforward approach to discover two actually linked users is to measure their similarity by comparing the common locations that they both have visited. However, users normally share few co-visited locations across different platforms due to the data sparsity and data imbalance problems.
To overcome the challenge, we propose a kernel density estimation (KDE) based method to accurately characterize the spatial and temporal patterns of
an individual's check-in activities and then perform user account linkage based on these patterns, inspired by \cite{lichman2014modeling}.
Although KDE is able to alleviate the data sparsity, this approach is inherently time consuming \cite{Wand1994Fast}\cite{lopez2015efficient}\cite{lopez2016kernel}, and the complexity is $O(m^2n^2)$ as presented in Theorem \ref{sec6:naive kde complexity}.

To improve the computational efficiency, we propose an index-based KDE. Specifically, we divide the space into $d\times d$ grid cells and $M$ time periods, and then each user is represented by a sequence of grid cells and time periods with corresponding probability. Compared with representing a user with a sequence of check-in records, the index based method is more efficient as the total number of grid cells and time periods is much smaller than that of check-ins. To further improve the efficiency, we propose a novel pruning strategy to reduce the number of account pairs to be measured, where only top-$k$ nearest neighbors of a user account are considered. Another benefit brought by index-based KDE is to relieve the data sparsity and imbalance problem. This is because although a user often posts different check-in activities on different social network platforms, the spatial distribution (e.g., the cell distribution) of his/her check-in records generated on each platform tends to be similar to each other \cite{Chen2017Exploiting}.

To improve the effectiveness, we firstly develop a density based method to delete outliers, where abnormal grid cells and time periods are pruned before the calculation of user account similarity. Additionally, we design an entropy-based weight scheme for locations and grid cells to reduce the impact of negative coincidence. As we introduced before, people tend to visit popular locations, which leads to the location coincidence. Obviously, these locations usually have large entropy in terms of the visited users. However, they are less useful in distinguishing users from each other. In contrast, private locations (e.g., homes, offices, etc.) are more discriminative in identifying users and usually have smaller entropy in terms of the visited users. Therefore, the key idea of our entropy-based weight calculation is to penalize locations and grid cells with high entropy by assigning low weights.

Apart from the evaluation of effectiveness of the proposed framework on several real-world datasets, some carefully generated synthetic/noisy datasets are also used to evaluate the scalability and robustness of HFUL. Additionally, we study the application of HFUL by user prediction, location prediction, and time prediction.

Different from most existing studies that mainly consider the effectiveness of user account linkage, our proposed framework HFUL has taken more factors into account. 1) \textbf{Efficiency.} The time cost of user account linkage is a significant factor for HFUL, since it means whether the framework can be widely used in many real applications. 2) \textbf{Effectiveness.} The effectiveness of user account linkage refers to the precision, recall, and F1. To achieve high effectiveness, we propose a novel kernel density estimation method for user account similarity calculation, and design a Renyi entropy-based weighting scheme to account for the importance of each extracted feature. 3) \textbf{Scalability.} The scalability, which is also an important factor that affects the availability of the proposed framework in real applications, refers to the stable performance of HFUL when there are a large-scale of user accounts to be measured and linked. 4) \textbf{Robustness.} The robustness refers to the reliable performance of HFUL while linking user accounts from datasets with many noisy records. 5) \textbf{Application.} Successfully linking users across different social platforms is able to benefit many real-world applications. In this paper, we show how to extend HFUL to several real applications including user prediction, time prediction, and location prediction.

\subsection{Our Contributions}
In this study, several approaches have been proposed to tackle the challenges that we are facing in user account linkage across location-aware social networks. To sum up, we make the following contributions.
\begin{itemize}
	\item We are the first to propose a hybrid framework to perform user account linkage with spatio-temporal data by considering effectiveness, efficiency, scalability, robustness, and application simultaneously.
	\item To tackle the data sparsity problem, we design a novel algorithm based on kernel density estimation. To tackle the data imbalance problem, we construct new index structures in both spatial and temporal domain. Furthermore, we design an entropy-based weight scheme with the goal of alleviating the challenges caused by negative coincidence.
	\item We reduce the computational complexity of the proposed framework by designing a novel pruning strategy to retrieve top-$k$ candidates for each user account.
	\item We have conducted extensive experiments, where the real-world datasets based results demonstrate the high performance of our proposed framework HFUL in terms of effectiveness, efficiency, and application; the synthetic datasets based results demonstrate the high robustness and scalability of HFUL.
\end{itemize}

Compared to the conference version of this study \cite{Chen2018Effective}, we make the following improvements:

\begin{itemize}
	\item Apart from spatial features, temporal features are also taken into account for better measuring the similarity between different user accounts. We characterize users from a comprehensive perspective by constructing a spatio-temporal index in Section \ref{sec8:framework optimization},  instead of characterizing them only based on spatial information.
	
	\item An outlier detection method is proposed to improve the effectiveness of user account linkage, where abnormal records are detected and removed before measuring user account similarity. A novel pruning strategy is developed to reduce the number of user accounts pairs to be measured from $m^2$ to $mk$, where $m$ and $k$ denote the number of user accounts on each platform and neighbors to be considered respectively. .                                 
	
	\item To study the applications of our work, we investigate the user prediction, location prediction, and time prediction on the cross-platform datasets by fusing check-ins belonging to the same user after linkage.
	\item Efficiency, effectiveness, scalability, robustness, and application of our proposed framework are investigated in experiments.
\end{itemize}

The rest of the paper is organized as follows. We present the related work in Section \ref{sec2:related work}, and formulate the problem in Section \ref{sec3:problem definition}. The overview of this work is presented in Section \ref{sec4:Overview}, and the KDE-based user similarity is introduced in Section \ref{sec5:proposed algorithm}. We construct the spatial and temporal index structures in Section \ref{sec6:index construction}, and optimize the proposed framework in Section \ref{sec8:framework optimization}, and investigate the application of HFUL in Section \ref{sec9:HFUL application}. The experiments are conducted in Section \ref{sec10:experiment study}, and the paper is concluded in Section \ref{sec11:conclusion}.


\section{Related Work}\label{sec2:related work}
The related studies, which contain the cross-platform user account linkage and kernel density estimation in spatio-temporal database, are discussed in this section.
\vspace{-15pt}
\subsection{Cross-platform User Linkage}
The increasing popularity of social networks has enabled more and more people to participate in multiple online services \cite{shu2017user}\cite{Huynh2020Adaptive}. Linking the same users across different platforms brings a great opportunity to fully understand users' behaviors and provide better recommendations \cite{Zhang2018Discrete}\cite{Wang2019UserIdentity}\cite{Chen2020Multi-level}. The study is firstly proposed in \cite{zafarani2009connecting}, where cross-community identities are connected with corresponding websites by measuring the identity similarity with usernames. Vosecky et al. \cite{vosecky2010user} proposed a method to identify users based on web profile matching and further extended its effectiveness by incorporating the user's friend network. To investigate whether users can be identified across systems based on their tag-based profiles, an aggregate profile was constructed by combining usernames and user tags \cite{iofciu2011identifying}. Following these studies, more abundant information was considered to link user accounts \cite{liu2013what}\cite{zafarani2013connecting}\cite{peled2013entity}\cite{liu2014hydra}\cite{shen2014controllable}. To build a comprehensive user profile for improving online services, various sources of complementary information were integrated \cite{zafarani2013connecting}, where the username features, prior-username features, and the relation between the candidate usernames
and prior usernames were taken into account. To match user accounts from different online social networks, Peled et al. \cite{peled2013entity} used supervised learning techniques to construct different classifiers, where three main types of features were utilized, i.e., name based features, social network topological based features, and general user info based features. To address the multi-platform user identity linkage problem, Mu et al. \cite{Mu2016User} proposed two effective algorithms, a batch model ULink and an online model ULink-On, based on latent user space modeling. Inspired by the fast development of the embedded technology, some people turn to investigate the cross-platform linkage with embedding \cite{Zhou2018Deeplink}\cite{Xie2018Unsupervised}\cite{Wang2019User}\cite{Liu2019ABNE}\cite{Zhou2019Translink}\cite{Fu2020Deep}. A novel framework called ``factoid embedding'' is proposed by \cite{Xie2018Unsupervised}, the core idea of the work is that each piece of information about a user identity describes the real identity owner, and thus distinguishes the owner from others. An attention-based network embedding model was proposed in \cite{Liu2019ABNE}, and the study contains two main components: a masked graph attention mechanism and an embedding algorithm which tries to learn a common vector space. Fu et al. \cite{Fu2020Deep} proposed a deep multi-granularity graph embedding model DeepMGGE, which utilizes the random walk to capture the higher-order structural proximities.

In recent advances \cite{Han2016Social}\cite{Riederer2016Lingking}\cite{Gao2018User}, researchers focused on using location data to achieve user account linkage. By utilizing the user-generated location data in social media platforms, a co-clustering-based framework was proposed \cite{Han2016Social}, where account clusterings in spatial and temporal and dimensions were carried out synchronously. To address the challenges in general cross-domain case, where users have different profiles independently generated from a common but unknown pattern, a generic and self-tunable algorithm that leverages any pair of sporadic location-aware datasets was proposed to determine the most likely matching between users \cite{Riederer2016Lingking}. To answer the question: is movement history sufficiently representative and distinctive to identify an individual? Jin et al. \cite{Jin2019Moving} formalized the problem of moving object linking as a $k$-nearest neighbor search on the collection of signatures, and aimed to improve efficiency considering the high dimensionality of signatures and the large cardinality of the candidate object set. To estimate two users whether have a social link, Zhang et al. \cite{Zhang2020Social} devised a novel multiview matching network MVMN, which contains three components, i.e., location matching module, time-series matching module, and relation matching module.

\subsection{Kernel Density Estimation}
Kernel density estimation (KDE) is a statistical technique for estimating a probability density function from a random sample set \cite{scott1985kernel}\cite{silverman1986density}. As a common tool, KDE has been explored in various areas for different purposes \cite{lopez2015efficient}, especially in spatio-temporal database \cite{zhang2013igslr}\cite{lichman2014modeling}\cite{hulden2015kernel}\cite{Backurs2019Space}\cite{Hohl2019Spatiotemporal}. 
To study the personalized geographical influence of locations on a user's behaviors, the kernel density estimation was used to model the personalized distribution of the distance between any pair of locations \cite{zhang2013igslr}. To understand urban human activity and mobility patterns, Hasan et al. \cite{hasan2013understanding} applied a two dimensional Gaussian kernel to estimate the check-in density of each grid cell. To investigate the spatio-temporal clustering of trajectories, a Gaussian kernel function was used to calculate the spatio-temporal kernel density of each trajectory unit \cite{zhang2014clustering}. To determine the geographical point of a text document, Hulden et al. \cite{hulden2015kernel} investigated an enhancement of common methods by kernel density estimation. To address the limitations of existing methods on understanding traffic accidents occur, a new method called Spatial-Temporal Network Kernel Density Estimation (STNKDE) is proposed in \cite{Romano2017Visualizing}. To examine whether a kernel density map could be reverse-transformed to its original map with discrete crime locations, Wang et al. \cite{Wang2019How} used the Epanecknikov, a default setting in ArcGIS for density mapping, to examine its impact on the deconvolution process. A new AE location method using tri-variate kernel density estimator was developed in \cite{Zhou2019Acoustic}, and the experimental results verified that the proposed method was more accurate and effective than traditional methods in the location performance. Coleman et al. \cite{Coleman2020Sub} proposed RACE, an efficient sketching algorithm for kernel density estimation on high-dimensional streaming data. The algorithm compresses a set of N high dimensional vectors into a small array of integer counters, and the array is sufficient to estimate the kernel density for a large class of kernels.

Connecting user accounts across different social platforms has been well studied by previous work from different perspectives, yet there is no study considers the effectiveness, efficiency, scalability, robustness, and application of user account linkage synchronously in spatio-temporal domain. Consequently, we propose the framework HFUL in this work to address the issue.


\section{Problem Definition}\label{sec3:problem definition}

In this section, we first present the notations that used throughout the paper in Table \ref{sec:sec1:tab:tab1} and then formulate the problem.
\vspace{-10pt}
\begin{table}[h]
	\centering
	\caption{Definitions of notations}\label{sec:sec1:tab:tab1}
	\begin{tabular}{l|l}
		\hline
		Notation 		& Definition \\ \hline
		$u$ 		    & A user account \\ \hline
		$U$ 		    & A set of user accounts \\ \hline
		$l$             & A location in the form of $(lat, lng)$\\ \hline
		$r_u$ 		    & A check-in record in the form of $(l, t)$\\ \hline
		$R$             & A set of check-in records \\ \hline
		$f(\cdot)$      & Probability density function \\ \hline
		$K_h(\cdot)$    & Gaussian kernel function \\ \hline
		$g$             & A grid cell \\ \hline
		$\varpi(g)$     & Probability of grid cell $g$ \\ \hline
		$\omega(g)$     & Weight of grid cell $g$ \\ \hline
		$G(u)$          & Grid representation of $u$ \\ \hline
		$T$             & A time period \\ \hline
		$\varpi(T)$     & Probability of time period $T$ \\ \hline
		$\omega(T)$     & Weight of time period $T$ \\ \hline
		$\Gamma(u)$     & Time period representation of $u$ \\ \hline
		$GT(u)$         & Spatio-temporal representation of $u$ \\ \hline
		$\mathcal{O}$   & A set of user account pair candidates \\ \hline
		$S_r(u_1,u_2)$  & Spatial similarity between $u_1$ and $u_2$ \\ \hline
		$S_t(u_1,u_2)$  & Temporal similarity between $u_1$ and $u_2$ \\ \hline
		$S_\Delta$      & Similarity threshold \\ \hline
		$S(u_1,u_2)$    & Final similarity between $u_1$ and $u_2$ \\\hline
	\end{tabular}
\vspace{-10pt}
\end{table}

On social networks, many users share ideas and check in after taking activities at a place. Then, the following information will be recorded and sent to the server: a unique user account id that distinguishes it from others; location information that consists of latitude and longitude; and time-stamp of the check-in record \cite{pham2013ebm}.

\begin{defn}\label{3:spatio-temporal record}
	\textbf{Check-in Record}. A check-in record of a user is defined as $r_u=(l, t)$, where $u$ is a user account, $l$ is defined as $(lat, lng)$ with $lat$ represents latitude and $lng$ represents longitude, and $t$ is the time-stamp.
\end{defn}

\begin{figure*}
	\centering
	\includegraphics[width=1.0\textwidth]{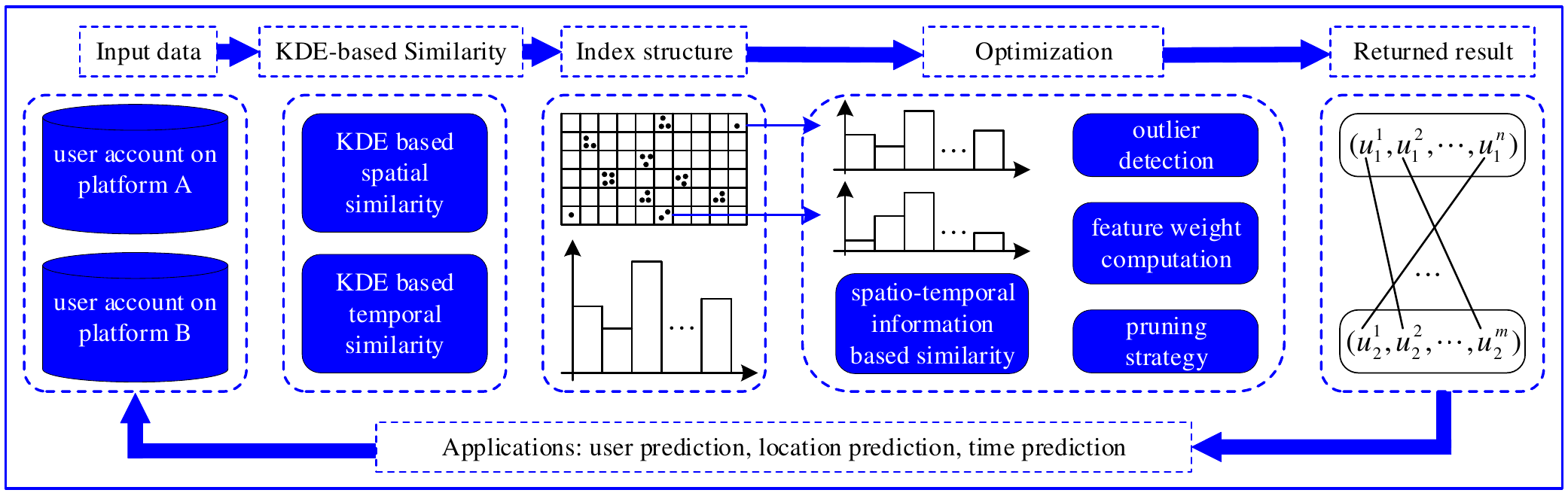}\\
	\caption{Overview of the proposed framework HFUL}
	\label{sec4:framework overview}
	\vspace{-10pt}
\end{figure*} 

Note that, the time-stamps can be used to distinguish records from each other. For instance, given two records $r_{u_1}=(l^1,t^1)$ and $r_{u_2}=(l^2,t^2)$, they are defined as different records if $t^1\neq t^2$ even though $u_1=u_2$ and $l^1=l^2$. This definition is appropriate, as a user may frequently check in  at the same place where he/she usually visits. The semantic information behind the records with the same location may be diverse. For example, a user may check in at a cafeteria on Monday, but check in at the same place on Sunday.

\textbf{Problem Formulation}. Given two user account sets $U_1=\{{u_1^1, u_1^2,\cdots,u_1^{m_1}}\}$ and $U_2=\{u_2^1, u_2^2,\cdots,u_2^{m_2}\}$ on two different location-aware social networks, where each user account is associated with a set of check-in records, our goal is to identify all account pairs $(u_1^i, u_2^j)$ of the same user from $\mathcal{O}=\{(u_1^i,u_2^j)|u_1^i\in U_1,u_2^j\in U_2\}$, on condition that $S(u_1^i, u_2^j)\geq S_\Delta$, where $S_\Delta$ is a given similarity threshold.

\section{Overview of HFUL}\label{sec4:Overview}
To link user accounts across different social platforms with location data, we propose the framework HFUL, which is composed of the following four main components as illustrated in Fig. \ref{sec4:framework overview}.

\textbf{KDE-based Similarity.} To tackle the problem ``data sparsity" introduced in Section \ref{1:introduction}, we propose a kernel density estimation-based algorithm, where the similarity between two user accounts is directly measured based on the naive KDE introduced in Section \ref{sec5:proposed algorithm}.

\textbf{Index Construction.} To reduce the high computational complexity brought by kernel density estimation, we design two indexes, i.e., grid map and time period structure, to organize the input data with grid and time representation.

\textbf{Optimization.} The proposed framework is optimized with the following steps: 1) instead of computing spatial and temporal similarity independently, we measure the user account similarity by considering these two parts of information simultaneously; 2) a density based clustering is developed to detect outliers; 3) an entropy based method is proposed to compute the weight for each grid cell and time period; 4) a novel algorithm is developed to retrieve candidates and an upper bound is proposed to further reduce the number of candidates to be measured. 

\textbf{Application of HFUL.} We obtain abundant data for each user following linkage, based on the cross-platform dataset we investigate the application of HFUL, i.e., user prediction, location prediction, and time prediction.


\section{KDE-based Similarity}\label{sec5:proposed algorithm}
In this section, we propose a kernel density estimation-based solution to perform user account linkage across location-aware social networks. The intuition behind our solution is that given two user accounts $u_1^i$ and $u_2^j$ of the same user on two different LBSNs, the distributions of her/his generated check-in records on the two LBSNs are similar to each other, even if the user posts different check-ins on these two platforms. For each user account pair $(u_1^i, u_2^j)$ in the Cartesian product $\mathcal{O}=U_1\times U_2=\{(u_1^i,u_2^j)|u_1^i\in U_1,u_2^j\in U_2\}$, we first compute their KDE-based similarity $S(u_1^i,u_2^j)$. Then, based on the inferred similarity and a user-defined similarity threshold $S_\Delta$, we decide whether these two accounts belong to the same user.

A straightforward way to measure the similarity between two user accounts  with discrete check-in records is to directly compare the records happened at same locations. Unfortunately, as discussed in Section \ref{1:introduction}, user generated check-in records on location-aware social networks are extremely sparse. Moreover, the issue of data missing worsens the situation. In light of these two challenges, we propose a kernel density estimation (KDE) based solution, inspired by its success in modeling individual-level location data~\cite{lichman2014modeling}. Kernel density estimation is a non-parametric method for estimating the probability density function of a sample set with unknown distribution. Given a set of locations $L=(l^1,l^2,\cdots,l^{n})$ and a location $l^\prime$, where each location is a two-dimensional tuple in the form of $(lat, lng)$,  the density of $l^\prime$ over $L$ is estimated as follows:
\begin{align}
	f(l^\prime|L,h)
	&=\dfrac{1}{n}\sum\limits_{i=1}^{n}K_h(l^\prime,l^i)\\
	K_h(l^\prime,l^i)\label{4:baseline kernel function}
	&=\dfrac{1}{2\pi h}\exp\big(-\dfrac{(l^\prime-l^i)^2}{2h^2}\big)
\end{align}
where $K(\cdot)$ is the Gaussian kernel function, $h$ is a bandwidth parameter, and $(l^\prime - l^i)$ is defined as the Euclidean distance between $l^\prime$ and $l^i$.

We use the probability density function $f(l^\prime|L,h)$ to denote the similarity between $L$ and $l^\prime$.
The similarity value  $f(l^\prime|L,h)$ will be large if $l^\prime$ is close to the points in $L$. In contrast, the value of $f(l^\prime|L,h)$ is small if points in $L$ are far away from $l^\prime$. 
Given two user accounts $u_1$ and $u_2$ with check-in record sets $R_{u_1}=(r_1^1,r_1^2,\cdots,r_1^{n_1})$ and $R_{u_2}=(r_2^1,r_2^2,\cdots,r_2^{n_2})$, the spatial similarity between $u_1$ and $u_2$ is defined as:
\begin{align}
	S_r(u_1,u_2)\label{4:S_r(u1,u2)}
	&=\dfrac{1}{n_1}\sum\limits_{i=1}^{n_1}f(r_1^i.l_1^i|R_{u_2},h)\\\nonumber
	&=\dfrac{1}{n_1n_2}\sum\limits_{i=1}^{n_1}\sum\limits_{j=1}^{n_2}K_h(r_1^i.l_1^i,r_2^j.l_2^j)
\end{align}
where $r_1^i.l_1^i$ denotes the location of the record $r_1^i$. Similarly, we define the temporal similarity $S_t(u_1,u_2)$ as follows by replacing location in Eq.(\ref{4:S_r(u1,u2)}) with time-stamp.
\begin{align}
	S_t(u_1,u_2)\label{4:S_t(u1,u2)}
	&=\dfrac{1}{n_1}\sum\limits_{i=1}^{n_1}f(r_1^i.t_1^i|R_{u_2},h)\\\nonumber
	&=\dfrac{1}{n_1n_2}\sum\limits_{i=1}^{n_1}\sum\limits_{j=1}^{n_2}K_h(r_1^i.t_1^i,r_2^j.t_2^j)
\end{align}
Then, the similarity between $u_1$ and $u_2$ is defined as:
\begin{align}\label{4:S(u_1,u_2)}
S(u_1,
&u_2)=S_r(u_1,u_2)+S_t(u_1,u_2)\\\nonumber
&=\dfrac{1}{n_1n_2}\sum\limits_{i=1}^{n_1}\sum\limits_{j=1}^{n_2}\big(K_h(r_1^i.l_1^i,r_2^j.l_2^j)+K_h(r_1^i.t_1^i,r_2^j.t_2^j)\big)
\end{align}


\section{Index Construction} \label{sec6:index construction}
Kernel density estimation is an important statistical technique in data analysis. According to Eq. (\ref{4:S(u_1,u_2)}), it requires $2n_1n_2$($|R_{u_1}|=n_1$ and $|R_{u_2}|=n_2$) kernel evaluations to measure the similarity $S(u_1,u_2)$, and the complexity of this method is $O(m^2n^2)$ as presented in Theorem \ref{sec6:naive kde complexity}. Obviously, the naive evaluation of KDE is very time consuming, especially for large-scale datasets with millions of check-ins. To speed up the evaluation of KDE, we propose novel index structures to organize the spatial and temporal data.

\begin{figure}[h]
	\centering
	\includegraphics[width=0.49\textwidth]{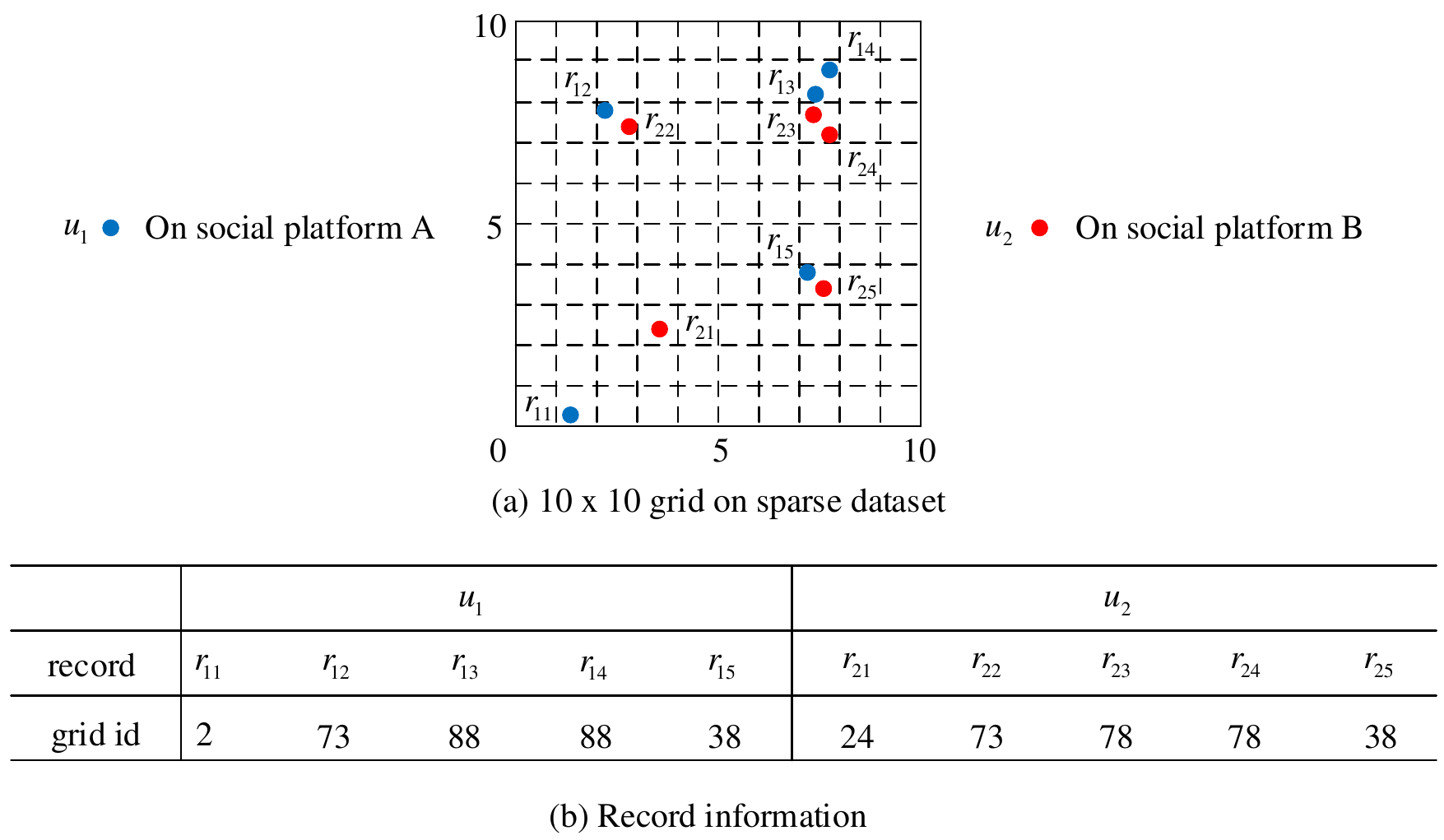}\\
	\caption{Grid structure}
	\label{5:grid division}
	\vspace{-10pt}
\end{figure}

\begin{figure}[h]
	\centering
	\includegraphics[width=0.4\textwidth]{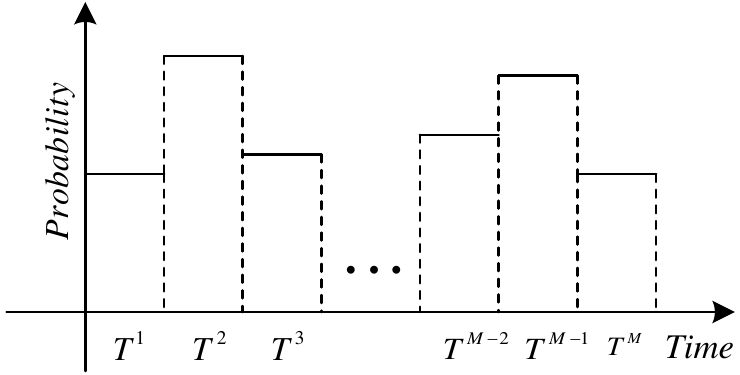}\\
	\caption{Time period structure}
	\label{5:time division}
	\vspace{-15pt}
\end{figure}

As shown in the example of Fig. \ref{5:grid division}(a), we divide the space into $10\times10$ square cells, and the grid id and record information are presented in Fig. \ref{5:grid division}(b). By assigning each cell a unique numerical id from bottom to top and from left to right, $u_1$ and $u_2$ are represented by a set of discrete cells, i.e., $\{2,73,88,38\}$ and $\{24,73,78,38\}$. Similarly, in temporal domain, we divide time into different periods. In Fig. \ref{5:time division}, the time is divided into $M$ periods $\{T^1,T^2,\cdots,T^M\}$. Then, each user can be represented by a set of time periods when he/she has visited. Due to the personal interests and geographical influence \cite{Yuan2013WhoWW}, the probabilities that a user visits different places are different in real life. Even though the same location, the user may visit at different time periods. As a result, we propose to compute the grid and time period probability for each user account based on Eq. (\ref{5:calculation of grid confidence}).

\vspace{-10pt}

\begin{defn}\label{5:grid confidence}
	\textbf{Grid and Time Period Probability}. Given a user account $u$ with a set of check-in records $R_u=(r^1,r^2,\cdots,r^n)$, the probability of each grid cell and time period visited by $u$ is defined as:
	\begin{align}\label{5:calculation of grid confidence}
		\varpi(g^i)=\dfrac{\Theta(g^i)}{|R_u|},~\varpi(T^i)=\dfrac{\Theta(T^i)}{|R_u|}
	\end{align}
\end{defn}
where $\Theta(g^i)$ denotes the number of records falling into the grid cell $g^i$, and $\Theta(T^i)$ represents the number of records falling into the time period $T^i$.

Based on the grid id and the calculated grid probability, the grid representation of each user account is defined as $G(u)=\{(g^1,\varpi(g^1)),(g^2,\varpi(g^2)),\cdots,(g^N,$ $\varpi(g^N))\}$. Continue the example in Fig. \ref{5:grid division}, we have $G(u_1)$ $=\{(2,0.2),(73,0.2),(88,0.4),(38,0.2)\}$ and $G(u_2)=\{(24,$ $0.2),(73,0.2),(78,0.4),(38,0.2)\}$. Based on the grid representation, we redefine the computation of KDE and similarity $S_r(u_1,u_2)$ as follows:
\begin{align}\label{5:grid base similarity}
&f(g_1^i|G(u_2),h)=\dfrac{1}{N_2}\sum\limits_{j=1}^{N_2}K_h(g_1^i,g_2^j)\\\nonumber
&K_h(g_1^i,g_2^j)=\dfrac{1}{2\pi h}\exp\big(-\dfrac{(g_1^i-g_2^j)^2}{2h^2}\big)\varpi(g_1^i)\varpi(g_2^j)\\\nonumber
&S_r(u_1,u_2)=\dfrac{1}{N_1N_2}\sum\limits_{i=1}^{N_1}\sum\limits_{j=1}^{N_2}K_h(g_1^i,g_2^j)
\end{align} where the grid representations of $u_1$ and $u_2$ are $G(u_1)=\{(g_1^1,\varpi(g_1^1)),(g_1^2,$ $\varpi(g_1^2)),\cdots,(g_1^{N_1},\varpi(g_1^{N_1}))\}$ and $G(u_2)$ $=\{(g_2^1,\varpi(g_2^1)),(g_2^2,\varpi(g_2^2)),\cdots,(g_2^{N_2},\varpi(g_2^{N_2}))\}$, respectively. $(g_1^i-g_2^j)$ denotes the Euclidean distance between the center coordinates of cells $g_1^i$ and $g_2^j$. Compared with the naive evaluation of KDE, the grid representation is a coarse-grained method, where the grid cell is the basic unit that may contain many points. Note that, implementing KDE with grid representation is able to: (1) reduce the number of kernel function evaluation, since we have $|G(u)|\leq|R_u|$ for each user account; (2) alleviate the data imbalance problem, since the probability that a user visits a specific geographical region tends to be similar across two LBSNs \cite{Chen2017Exploiting}. Grid-based kernel density estimation is more efficient than the method in Section \ref{sec5:proposed algorithm}, especially given a large dataset.

Similarly, in temporal domain, we can redefined the computation of Gaussian kernel function $K(\cdot)$ and similarity $S_t(u_1,u_2)$ by representing each user account with a sequence of time periods with corresponding probability as follows:
\begin{align}\label{5:time period base similarity}
&f(T_1^i|\Gamma(u_2),h)=\dfrac{1}{M_2}\sum\limits_{j=1}^{M_2}K_h(T_1^i,T_2^j)\\\nonumber
&K_h(T_1^i,T_2^j)=\dfrac{1}{2\pi h}\exp\big(-\dfrac{(T_1^i-T_2^j)^2}{2h^2}\big)\varpi(T_1^i)\varpi(T_2^j)\\\nonumber
&S_t(u_1,u_2)=\dfrac{1}{M_1M_2}\sum\limits_{i=1}^{M_1}\sum\limits_{j=1}^{M_2}K_h(T_1^i,T_2^j)
\end{align} where the time period representations of $u_1$ and $u_2$ are assumed as $\Gamma(u_1)=\{(T_1^1,\varpi(T_1^1)),(T_1^2,\varpi(T_1^2)),$ $\cdots,$ $(T_1^{M_1},\varpi(T_1^{M_1}))\}$ and $\Gamma(u_2)=\{(T_2^1,\varpi(T_2^1)),(T_2^2,\varpi(T_2^2$ $)),\cdots,$ $(T_2^{M_2},\varpi(T_2^{M_2}))\}$, respectively. $M_1$ and $M_2$ denote the number of time periods containing at least one check-in of $u_1$ and $u_2$ respectively, and $(T_1^i-T_2^j)$ denotes the Euclidean distance between the centers of periods $T_1^i$ and $T_2^j$. By constructing the time period structure, we can also improve the efficiency and alleviate the data imbalance problem in temporal domain. Following the redefinition of $S_r(u_1,u_2)$ and $S_t(u_1,u_2)$, the similarity $S(u_1,u_2)$ is redefined as:
\begin{align}\label{5: user final similarity}
&S(u_1,u_2)=S_r(u_1,u_2) + S_t(u_1,u_2)\\\nonumber
&=\dfrac{1}{N_1N_2}\sum\limits_{i=1}^{N_1}\sum\limits_{j=1}^{N_2}K_h(g_1^i,g_2^j)+\dfrac{1}{M_1M_2}\sum\limits_{i=1}^{M_1}\sum\limits_{j=1}^{M_2}K_h(T_1^i,T_2^j)
\end{align}

\section{Optimization of Framework}\label{sec8:framework optimization}
By constructing the index structure, we can improve the efficiency of user account linkage. This is because the number of grids and time periods is less than that of check-in records. Next, we use the following outlier detection and pruning strategy to optimize the effectiveness of HFUL.

\subsection{Spatio-temporal information based comprehensive user account similarity}\label{sec8:comprehensive index}
To find the actually linked user accounts across different social platforms, the aforementioned methods calculate the spatial similarity $S_r(u_1,u_2)$ and temporal similarity $S_t(u_1,u_2)$ based on location and time information respectively. To improve the efficiency, the grid and time period structures are developed to organize the input data with corresponding representations. However, in real life, a check-in record is by default a spatio-temporal event. The location $(lat,lng)$ and time $t$ of a check-in record are usually sent to the server synchronously if a user decides to share his/her status associated with an address in real life. Consequently, to exploit users' behaviors more precisely and achieve higher performance (i.e., precision, recall, and F1), we consider the spatial and temporal information simultaneously during the measurement of user account similarity.

\begin{figure}[h]
	\centering
	\includegraphics[width=0.49\textwidth]{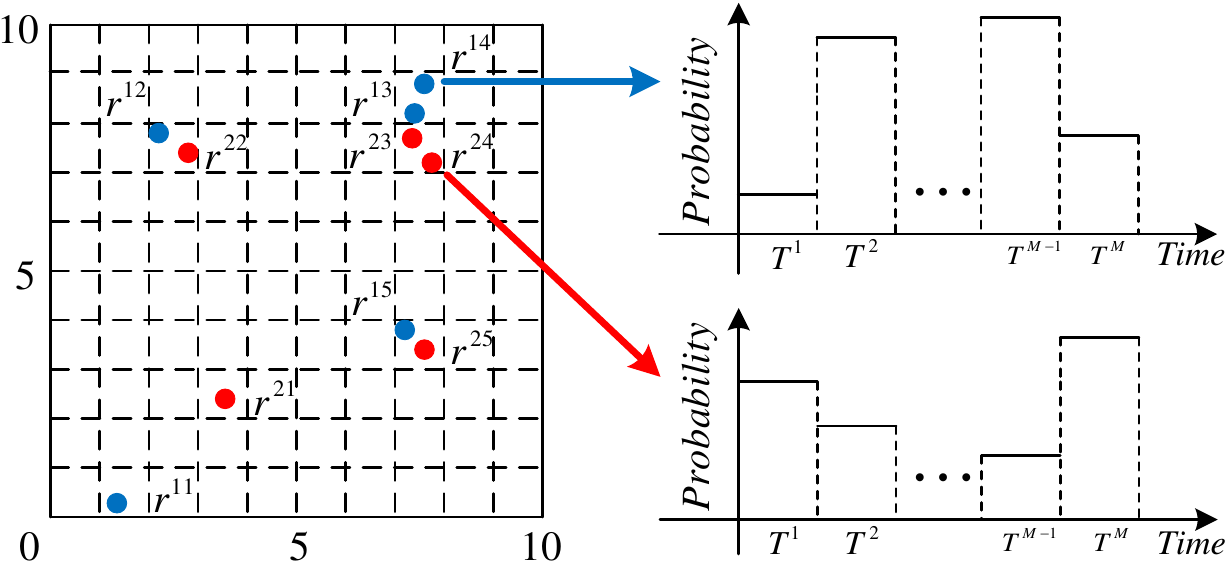}\\
	\caption{Spatio-temporal index}
	\label{sec7:time distribution}
	\vspace{-20pt}
\end{figure}

Continue the example in Fig. \ref{5:grid division}, instead of mapping the check-in records of a user into grid cells and time periods independently, each user is characterized from a joint spatio-temporal perspective. As shown in Fig. \ref{sec7:time distribution}, we extract the time distribution of a user in each grid cell. Given a set of check-in records $R_u=(r^1,r^2,\cdots,r^n)$ of $u$, the spatio-temporal representation $GT(u)$ of a user account is defined as:
\begin{align}\label{sec8:grid time probability}
&\varpi(T_{g^i}^j)= \dfrac{\Theta(T_{g^i}^j)}{|R_{u,g_i}|}\\ \nonumber
&\Gamma(g^i)=\{(T_{g^i}^1,\varpi(T_{g^i}^1)),\cdots,(T_{g^i}^{M_{g^i}},\varpi(T_{g^i}^{M_{g^i}}))\}\\\nonumber
&GT(u)=\{(g^1,\varpi(g^1),\Gamma(g^1)),\cdots,(g^N,\varpi(g^N),\Gamma(g^N))\}
\end{align}
where $|R_{u,g_i}|$ is the number of check-ins of $u$ in $g_i$, $\Theta(T_{g^i}^j)$ and $M_{g^i}$ denotes the number of records falling into the period $T^j$ and periods containing at least one record in $g_i$ respectively. Then, the Gaussian kernel function is redefined as:
\begin{align}\nonumber
&K_h(g_1^x,g_2^y)=\big[\dfrac{1}{2\pi h}\exp\big(-\dfrac{(g_1^x-g_2^y)^2}{2h^2}\big)\varpi(g_1^x)\varpi(g_2^y)\big]^\alpha\cdot\\\nonumber
&\big[\sum_{i=1}^{M_{g_1^x}}\sum_{j=1}^{M_{g_2^y}}\dfrac{1}{2\pi h}\exp\big(-\dfrac{(T_{g_1^x}^i-T_{g_2^y}^j)^2}{2h^2}\big)\varpi(T_{g_1^x}^i)\varpi(T_{g_2^y}^j)\big]^{1-\alpha}
\end{align}
where the parameter $\alpha(0\leq \alpha \leq 1)$ is a trade-off between the spatial and temporal information, which is obtained by maximizing the F1 score in experiments. $\varpi(T_{g_1^x}^i)$ denotes the probability of time period $T^i$ in grid $g_1^x$. Next, we have the new similarity $S(u_1,u_2)$ by replacing the Gaussian kernel function in Eq. (\ref{5:grid base similarity}). Compared with the method in Section \ref{sec6:index construction}, where the minimum granularities are $(g^i,\varpi(g^i))$ and $(T^i,\varpi(T^i))$ in spatial and temporal domain respectively, the comprehensive granularity $(g^i,\varpi(g^i),\Gamma(g^i))$ based approach is more likely to link user accounts precisely.

\vspace{-10pt}

\subsection{Outlier Detection}\label{sec6:outlier detection}
Due to the instability of GPS-enabled devices, there usually exist some outliers in spatio-temporal data. As these outliers are negative factors for precise linkage, we develop the following approach to detect and remove anomalous grids before computing similarity.

\begin{figure}[h]
	\centering
	\includegraphics[width=0.49\textwidth]{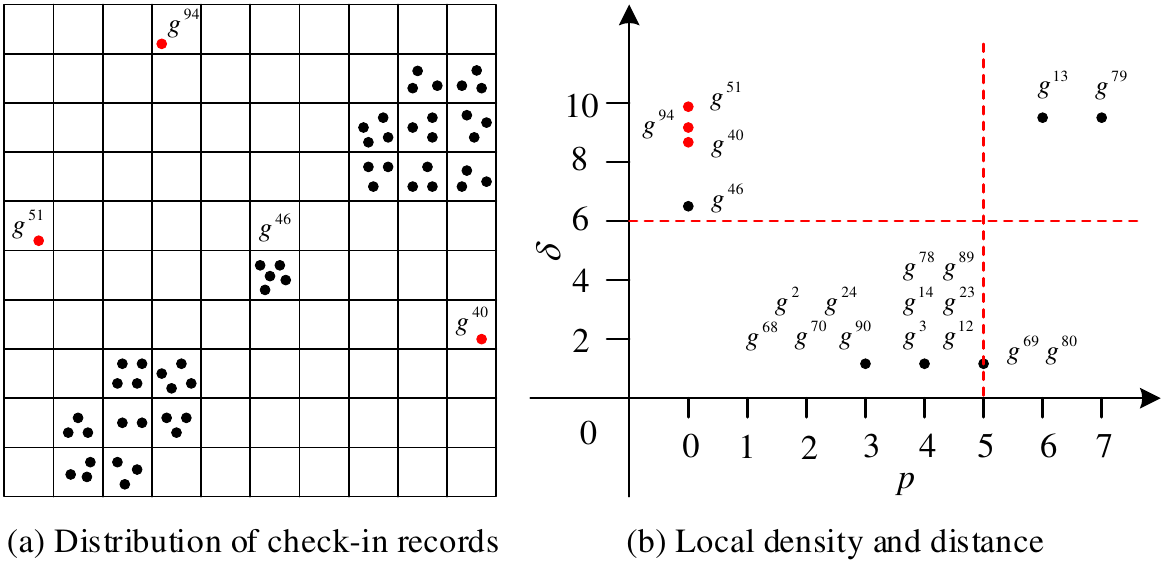}\\
	\caption{An example of outlier detection}
	\label{6:example of outlier}
\end{figure}

We design a novel method based on an advanced density-based clustering (DP) \cite{rodriguez2014clustering}, and the reason behind choosing to augment DP for detecting outliers is twofold. 1) There usually exist some outliers on real spatio-temporal datasets due to the instability of GPS-enabled devices \cite{Albanese2014Rough}\cite{Duggimpudi2019Spatio}, and DP has been proved to be able to ignore anomalous points \cite{rodriguez2014clustering}. 2) Many clustering algorithms require the user to set many parameters, while only two parameters are necessary for the DP algorithm \cite{Begum2015Accelerating}. The idea of DP is: cluster centers are surrounded by neighbors with lower local density and that they are at a relative large distance from any points with a higher local density. Given a user account $u$ with spatio-temporal representation $GT(u)=\{(g^1,\varpi(g^1),\Gamma(g^1)),\cdots,(g^N,\varpi(g^N),$ $\Gamma(g^N))\}$, we firstly compute the local density $p_{g^i}$ and the distance from grids with higher local density $\delta_{g^i}$ for each grid cell with following equations,
\begin{align}\nonumber
p_{g^i} 
& = \sum_{j}\chi(d_{g^i,g^j}-d_c), \left\{{\begin{array}{l}\chi(x)=1,~if~x<0\\\chi(x)=0,~ otherwise\end{array}}  \right.\\
\delta_{g^i}\label{6:grid delta}
& = \left\{{\begin{array}{l}\mathop{min}\limits_{{p_{g^j}}>p_{g^i}}(d_{g^i,g^j}),~if~p_{g^j}>p_{g^i}\\\mathop{max}\limits_j(d_{g^i,g^j}),~otherwise
	\end{array}}  \right.
\end{align}
where $d_c$ is the cutoff distance given by users. $p_{g^i}$ is the number of grids that are closer to the candidate grid $g^i$ than $d_c$. $\delta_{g^i}$ denotes the minimum distance between $g^i$ and any other grid with higher density, especially, $\delta_{g_i}$ is measured as the maximum distance $d_{g^i,g^j}$ if $g^i$ has the highest density. Following the calculation of $p_{g^i}$ and $\delta_{g^i}$, the top-$k$ centers with the highest value $\xi_{g^i}=p_{g^i}\times\delta_{g^i}$ are returned as cluster centers, and we assign different labels to these centers. Then, we assign each remaining grid to the same cluster as its nearest neighbor of higher density. If there exists a grid $g^i$ that do not belong to any cluster, then $g^i$ is defined as an outlier. However, in real life, there may exist some grid cells that contain many check-ins but without any neighbor, such as $g^{46}$ in Fig. \ref{6:example of outlier}. To avoid misjudgment, we need to set a probability threshold $\varpi_{_\Delta}$, i.e., if $\varpi(g)\geq \varpi_{_\Delta}$, then the grid cell $g$ is not an outlier even though it has no neighbor.

The main information to be used during the detection of outliers are locations of users, this is because the density of them is usually higher than that of time on location-based social networks. For example, in real life many users may share statues at same places in different time periods. In this case, the time information is useless to characterize a user, especially when the distribution of the user's records is very sparse. This significant phenomenon, i.e., the representativeness of spatial is usually much higher than that of time on location-based social networks, has been fully investigated and demonstrated in our experiments. 

Consider the example in Fig. \ref{6:example of outlier}, by computing the local density $p_{g^i}$ and distance $\delta_{g^i}$ for each grid and setting $\xi_{g^i}=5\times 6$, $\varpi_{_\Delta}=0.1$,  then we can delete anomalous grids $g^{40}$, $g^{51}$, and $g^{94}$. Even though these grids have large distance $\delta_{g^i}$, the local density of them is 0, as there is no record falling into their neighbor grids. Additionally, the probability of each of them is 0.025 based on Eq. (\ref{5:calculation of grid confidence}), which is less than $\varpi_{_\Delta}$. Note that, the grid $g^{46}$ with probability 0.125 cannot be deleted since $\varpi(g^{46})> \varpi_{_\Delta}$. Observed from this example, the novel method is able to detect outliers, and we can achieve higher precision through outlier detection, since the remaining grids are more likely to reflect the real behaviors of a user in real life. Next, we redefine the similarity $S(u_1,u_2)$ as follows:
\begin{align}\label{6:outlier user similarity}
&S(u_1, u_2) = \dfrac{1}{XY}\sum\limits_{x=1}^{X}\sum\limits_{y=1}^{Y}K_h(g_1^x,g_2^y)\\\nonumber
&=\dfrac{1}{XY}\sum\limits_{x=1}^{X}\sum\limits_{y=1}^{Y}\big\{[\dfrac{1}{2\pi\nonumber h}\exp\big(-\dfrac{(g_1^x-g_2^y)^2}{2h^2}\big)\varpi(g_1^x)\varpi(g_2^y)]^\alpha\cdot\\\nonumber
&[\sum_{i=1}^{M_{g_1^x}}\sum_{j=1}^{M_{g_2^y}}\dfrac{1}{2\pi h}\exp\big(-\dfrac{(T_{g_1^x}^i-T_{g_2^y}^j)^2}{2h^2}\big)\varpi(T_{g_1^x}^i)\varpi(T_{g_2^y}^j)]^{1-\alpha}\big\}
\end{align}
where $X$ and $Y$ denote the number of grids of $u_1$ and $u_2$ respectively after outlier detection.

\subsection{Feature Weight Computation}\label{sec8:feature weight}
Intuitively, the popular places, such as shopping mall, cafeteria, and cinema, are more attractive and more likely to be visited by many people than personal private places, such as home and office. This phenomenon leads to the low peculiarity of popular places, i.e., these places are useless for distinguishing users from each other. In contrast, the personal private places visited by less people are more discriminative \cite{Chen2017Exploiting}. In other words, the importance of different places are different. To achieve user account linkage with higher accuracy, we highlight discriminative grids and time periods with large weight, and lighten popular ones with small weight. Inspired by~\cite{pham2013ebm}, we propose to use Entropy from information theory to compute the weight of each grid cell and time period in this section.

Renyi entropy is a generalized version of Shannon entropy and it's defined as follows in our application:
\begin{align}\label{6:Renyi grid cell entropy}
&H(g) = \dfrac{1}{1-q}\log\sum\limits_{i=1}^{\mathcal{N}}\big(\dfrac{\Theta_{u_i}(g)}{|R_{u_i}|}\big)^q\\
&H(T) = \label{6:Renyi time period entropy} \dfrac{1}{1-q}\log\sum\limits_{i=1}^{\mathcal{M}}\big(\dfrac{\Theta_{u_i}(T)}{|R_{u_i}|}\big)^q
\end{align}where $\mathcal{N}$ and $\mathcal{M}$ denote the number of user accounts having check-ins in grid cell $g$ and time period $T$ respectively. Compared with Shannon entropy, the adjustable $q$ makes Renyi entropy much more expressive and flexible. Following the study in \cite{pham2013ebm}, the parameter $q$ indicates entropy's sensitivity to the number $\Theta_{u_i}(g)$ and $\Theta_{u_i}(T)$. Specifically, it has following properties:
\begin{itemize}
	\item If $q>1$, the entropy rewards the higher value of $\Theta_{u_i}(g)$ and $\Theta_{u_i}(T)$.
	\item If $q<1$, the entropy penalizes the higher value of $\Theta_{u_i}(g)$ and $\Theta_{u_i}(T)$.
	\item If $q=1$, it is the meeting point of Shannon entropy and Renyi entropy.
\end{itemize}

According to the Renyi entropy, we give the definitions of the grid weight and time period weight. 
\begin{align}
&\omega(g)=\label{sec8:grid cell weight} \exp\big(-H(g)\big)=\Big(\sum\limits_{i=1}^{\mathcal{N}}\big(\dfrac{\Theta_{u_i}(g)}{|R_{u_i}|}\big)^q\Big)^{\dfrac{1}{q-1}}\\
&\omega(T)= \label{sec8:time period weight} \exp\big(-H(T)\big)=\Big(\sum\limits_{i=1}^{\mathcal{M}}\big(\dfrac{\Theta_{u_i}(T)}{|R_{u_i}|}\big)^q\Big)^{\dfrac{1}{q-1}}
\end{align} To normalize the weight of grid and time period, we use $\omega(g^i)=\dfrac{\omega(g^i)}{\max~ \omega(g)}$ and $\omega(T^i)=\dfrac{\omega(T^i)}{\max~ \omega(T)}$, where $\max~ \omega(g)$ and $\max~ \omega(T)$ denote the maximum weight of all grids and time periods respectively. Next, we can update the Gaussian kernel function $K(\cdot)$ as follows:

\begin{small}
\begin{align}
&K_h(g_1^x,g_2^y)\label{sec8: final kernel}=\\
&\big[\dfrac{1}{2\pi\nonumber h}\exp\big(-\dfrac{(g_1^x-g_2^y)^2}{2h^2}\big)\varpi(g_1^x)\varpi(g_2^y)\omega(g_1^x)\omega(g_2^y)\big]^\alpha\cdot\big[\sum_{i=1}^{M_{g_1^x}}\sum_{j=1}^{M_{g_2^y}}\\\nonumber
&\dfrac{1}{2\pi h}\exp\big(-\dfrac{(T_{g_1^x}^i-T_{g_2^y}^j)^2}{2h^2}\big)\varpi(T_{g_1^x}^i)\varpi(T_{g_2^y}^j)\omega(T_{g_1^x}^i)\omega(T_{g_2^y}^j)\big]^{1-\alpha}
\end{align}
\end{small}

\noindent Then, we can obtain the final user account similarity $S(u_1,u_2)$ based on Eq. (\ref{sec8:final similarity}).
\begin{table*}
	\begin{align}\label{sec8:final similarity}
	&S(u_1, u_2)=\dfrac{1}{XY}\sum\limits_{x=1}^{X}\sum\limits_{y=1}^{Y}K_h(g_1^x,g_2^y)\\
	&=\dfrac{1}{XY}\sum\limits_{x=1}^{X}\sum\limits_{y=1}^{Y}\big[\dfrac{1}{2\pi \nonumber h}e^{-\dfrac{(g_1^x-g_2^y)^2}{2h^2}}\varpi(g_1^x)\varpi(g_2^y)\omega(g_1^x)\omega(g_2^y)\big]^\alpha\big[\sum_{i=1}^{M_{g_1^x}}\sum_{j=1}^{M_{g_2^y}}
	\dfrac{1}{2\pi h}e^{-\dfrac{(T_{g_1^x}^i-T_{g_2^y}^j)^2}{2h^2}}\varpi(T_{g_1^x}^i)\varpi(T_{g_2^y}^j)\omega(T_{g_1^x}^i)\omega(T_{g_2^y}^j)\big]^{1-\alpha}
	\end{align}
\end{table*}

\subsection{Grid-based Pruning Strategy}
To further improve the efficiency of HFUL, we propose the following strategy to reduce the complexity of user account pair similarity calculation.

As presented in Algorithm \ref{sec6:candidate retrieve}, we use the spatial information to retrieve user account pair candidates. The core idea of the pruning strategy is that: two user accounts have more common grid cells, then they are more likely to be an actually linked pair. During the process, an empty list $L_{u_1^i}$ is created to store the nearest neighbors of user account $u_1^i$, during each process. Then, for each grid cell $g^l$ in $G(u_1^i)$, we add the user account $u_2^j$ into $L_{u_1^i}$ if $u_2^j$ also has check-in records falling into $g^l$ (i.e., $\varpi_{u_2^j}(g^l)>0$). Next, we sort users accounts in $L_{u_1^i}$ to select the top-$k$ nearest neighbors of $u_1^i$ based on $|GT(u_1^i)\cap GT(u_2^j)|$, which is defined as:
\begin{align}\label{sec8:common grid}
&|GT(u_1^i)\cap GT(u_2^j)|\\\nonumber
&=\sum_{g\in GT(u_1^i),g\in GT(u_2^j)}min(\varpi_{u_1^i}(g),\varpi_{u_2^j}(g))
\end{align}
Finally, we add the candidate $(u_1^i,u_2^{j^\prime})$, which is likely to be an actually linked pair, into $\mathcal{O}$ by selecting the top-$k$ accounts in $\mathcal{L}_{u_1^i}$ (i.e., $\mathcal{L}_{u_1^i}[1:k]$). 

\begin{algorithm}
	\SetAlgoLined
	\caption{Candidate Retrieval}
	\label{sec6:candidate retrieve}
	\KwData {two user account sets $U_1=\{{u_1^1, u_1^2,\cdots, u_1^{m_1}}\}$, $U_2=\{{u_2^1, u_2^2, \cdots,u_2^{m_2}}\}$}
	\KwResult{ candidate set $\mathcal{O}$}
	\For{each user account $u_1^i$ in $U_1$}
	{
		create an empty list $\mathcal{L}_{u_1^i}$;\\
		\For{each grid cell $g^l$ in $GT(u_1^i)$}
		{
			add a user account $u_2^j$ into $\mathcal{L}_{u_1^i}$ if $u_2^j$ has $\varpi_{u_2^j}(g^l)>0$;
		}
		sort user accounts in $\mathcal{L}_{u_1^i}$ based on the value of $|GT(u_1^i)\cap GT(u_2^j)|$;\\
		$\mathcal{L}_{u_1^i}[1:k]\leftarrow$ top-$k$ user accounts in $\mathcal{L}_{u_1^i}$; \\
		\For{each user account $u_2^{j^\prime}$ in $\mathcal{L}_{u_1^i}[1:k]$}
		{
			add the user account pair candidate $(u_1^i,u_2^{j^\prime})$ into $\mathcal{O}$;
		}
	}
	\Return the candidate set $\mathcal{O}$
\end{algorithm}

For each candidate $(u_1^i,u_2^j)$ in the set $\mathcal{O}$, we can obtain the final similarity $S(u_1^i,u_2^j)$ based on Eq. (\ref{sec8:final similarity}) and return the pair on condition that $S(u_1^i,u_2^j)\geq S_\Delta$.

\subsection{Complexity Analysis}\label{sec6:complexity analysis}
The computational complexity is a significant factor affecting the efficiency and scalability of HFUL, thus we reduce the complexity of it by developing the above-mentioned pruning strategy.
\begin{thm}\label{sec6:naive kde complexity}
	The complexity of naive KDE-based method is $O(m^2n^2)$.
\end{thm}
\begin{proof}
	The similarity between two user accounts calculated based on the naive KDE-based method is presented in Eq. (\ref{4:S(u_1,u_2)}), and the complexity is $O(n^2)$ if both $u_1$ and $u_2$ have $n$ check-in records. Then, the final complexity of this method is $O(m^2n^2)$ if there exist $m$ user accounts on two given platforms respectively.
\end{proof}

\begin{thm}\label{sec6:complexity of two}
	The complexity of measuring the similarity between two user accounts based on Eq. (\ref{sec8:final similarity}) is $O(n^2)$.
\end{thm}
\begin{proof}
	In Eq. (\ref{sec8:final similarity}), the upper bound of $XYM_{g_1^x}M_{g_2^y}$ is $n^2$ on condition that each time period in grid cells visited by $u_1$ and $u_2$ at most contains one check-in, i.e., we need to compute the similarity between any two check-ins in this extreme case. Consequently, the complexity is $O(n^2)$.
\end{proof}

\begin{thm}\label{sec6:upper bound number}
	The complexity of measuring user account similarity based on the candidate set $\mathcal{O}$ obtained in Algorithm \ref{sec6:candidate retrieve} is $O(mkn^2)$.
\end{thm}
\begin{proof}
	In Algorithm \ref{sec6:candidate retrieve}, we only consider top-$k$ neighbors for each user account in $U_1$, thus the complexity of this method is $O(mkn^2)$.
\end{proof}

The efficiency of HFUL can be significantly improved with the pruning strategy, and the reason is twofold. 1) The upper bound of user account pairs to be considered has been reduced from $m^2$ to $mk$ based on the Algorithm \ref{sec6:candidate retrieve}, where the parameter $k$ is a constant with $k<<m$. 2) Although the upper bound of $XYM_{g_1^x}M_{g_2^y}$ is $n^2$, we have  $XYM_{g_1^x}M_{g_2^y}<n^2$ for most user account pairs during the calculation of similarity. This is because there are many records falling into the same time period in a grid cell, and the extreme case mentioned in Theorem \ref{sec6:complexity of two} is uncommon in real life. Additionally, we need some time to find top-$k$ neighbors from $\mathcal{L}_{u_i}$ in Algorithm \ref{sec6:candidate retrieve}, but this time is low-impact on the efficiency of our proposed method, since it is much smaller than the time saved from pruning search space. The experimental results in Section \ref{sec10:experiment study} also demonstrate this improvement.


\section{Application of HFUL}\label{sec9:HFUL application}
Intuitively, the goal of cross-platform user account linkage is to integrate the sources of complementary information from different platforms for each user. Obviously, each user can obtain more history data following the linkage. Based on these data, we can exploit users' behaviors more precisely. This is because compared with behavior analysis focusing on a specific platform, we have more insight into user behaviors with cross-platform datasets. The reason also answers the question ``Why to link accounts belonging to the same user across different platforms?''. Next, we investigate the application of HFUL on user prediction, location prediction, and time prediction. 

An individual's mobility usually centers at different personal geographical regions, historical check-in records of a user are usually generated at some specific regions, such as home region and 
work region \cite{Yuan2013WhoWW}. Based on the density-based clustering method (DP) mentioned in Section \ref{sec6:outlier detection}, we can extract the multinomial distribution of region $p(l|u)$ for each user. Obviously, grid cells falling into these regions usually have higher density compared with grid cells outside these regions. Next, we extract the multinomial distribution of time $p(t|u,l)$ in each region $l$ for all users, based on the temporal index introduced in  Fig. \ref{5:time division}. Following the analysis of users' historical behaviors, we investigate the following predictions.

\begin{itemize}
	\item \textbf{User Prediction.} Based on the region distribution $p(l|u)$ and time distribution $p(t|u,l)$, we can predict the likelihood of a user visiting a target region at a specific time $p(u|l,t)$. This could be very useful for merchants for planning purpose, or for them to target on specific costumers \cite{Yuan2013WhoWW}. Given location $(lat,lng)$ and time $t$, we firstly locate the region $l$ that the point $(lat,lng)$ falls into, then rank candidate users by $p(u|l,t)$, which is computed as follows:
	\begin{align}
		p(u|l,t)=\dfrac{p(u,l,t)}{\sum_{u^\prime} p(u^\prime,l,t)}
	\end{align}
	where $p(u,l,t)=p(u)p(l|u)p(t|u,l)$, and $p(u)$ denotes the percentage of records of $u$ in all records, i.e., $p(u)=|R_u|/\sum_{i=1}^{n}|R_{u_i}|$.
	
	\item \textbf{Location Prediction for User.} This task is to predict the region where a user stays at a given time. This would be useful for location-aware advertisement recommendation, and for people to arrange a meeting with a specific user or a group of users. Given a user $u$ and time $t$, we aim to rank all candidate region based on $p(l|u,t)$ with following method:
	\begin{align}
	p(l|u,t)=\dfrac{p(u,l,t)}{\sum_{l^\prime} p(u,l^\prime,t)}
	\end{align}
	
	\item \textbf{Time Prediction for User.} This task is to predict the time when a user may visit a specific region. This would be useful for time-based advertisement delivery and real-time recommendation. Formally, given a user $u$ and a location $(lat,lng)$, we can directly return the time interval $t$ with the maximum $p(t|u,l)$, since we have extracted the time distribution in the region $l$ that contains the location $(lat,lng)$.
\end{itemize}

During the user prediction and time prediction, there may exist some locations that cannot be contained by a historical region, we will select a nearest region to the given location for prediction in this case. 


\section{Experiment Study}\label{sec10:experiment study}
Extensive experiments are conducted in this section. First, we describe the experiment setup, which contains dataset introduction, baseline algorithms presentation, and evaluation metric discussion. Then, the effectiveness, efficiency, scalability, robustness, and application of the proposed framework HFUL are reported.
\vspace{-5pt}
\subsection{Experiment Datasets}
\textbf{Foursquare-Twitter (FTW)}. Foursquare and Twitter are two widely used social networks, where users can post statuses associated with location information. To investigate the performance of the proposed approach in linking cross-platform user accounts, we use the dataset provided by \cite{zhang2014transferring}\cite{Riederer2016Lingking}, where they select users with records presented in both platforms. The dataset contains 862 users with 13177 Foursquare records and 174618 Twitter records.

\textbf{Instagram-Twitter (ITW)}. Instagram is another popular photo-sharing application, where users can share pictures and videos with location information with mobile, desktop, laptop, and tablet. To link the user accounts across Instagram and Twitter with location data, we use the dataset processed by \cite{Riederer2016Lingking}. Similarly, each user of the dataset has check-in records generated in both platforms. The dataset contains 1717 users with 337934 Instagram records and 447366 Twitter records.

\textbf{GOW}. Gowalla is a location-based social network, where users can share their locations by checking-in. The dataset collected from Gowalla contains 1855 user accounts with 2097885 check-in records. To simulate user account linkage across two platforms, we randomly divide the dataset into two components GOWA and GOWB, i.e., the records of each user is randomly divided into two subsets with roughly equal size. 

\subsection{Compared Methods.} 
We compare the performance of our method with several state-of-the-art location based user account linkage approaches. Although existing methods \cite{zafarani2013connecting}\cite{liu2014hydra}\cite{Mu2016User} also work on user account linkage, their results are not comparable here, as they use different input data, such as text messages, user profile, and language style. Algorithms proposed by \cite{Jin2019Moving}\cite{Zhang2020Social} only make sense in trajectory data and cannot be extended to discrete check-in records.

\textbf{GRID}: The first method is based on of the work proposed by \cite{li2010mining}, where the top-$p\%=15\%$ grid cells with the maximum density are returned to denote a user. Based on these grid cells, we use following method to measure the similarity between a user account pair.
\begin{equation}\nonumber
S(u_1,u_2)=\sum\limits_{r_{1i}\in R_1}^{m}\sum\limits_{r_{2j}\in R_2}^{n}\dfrac{|r_{1i}\cap r_{2j}|}{|r_{1i}\cup r_{2j}|}\cdot min(f(r_{1i}),f(r_{2j}))
\end{equation}
where $f(r_{ij})$ denotes the density of the $j$-th grid of $u_i$.

\textbf{BIN}: The second method is proposed by \cite{Riederer2016Lingking}, where each record in region $l$ during time interval $t$ is associated with bin $(l,t)$. The similarity between $u_1$ and $u_2$ is defined as:
\begin{equation}\nonumber
S(u_1,u_2)=\sum\limits_{t\in T}\sum\limits_{l\in L}S(u_1,u_2,l,t)
\end{equation}
and the similarity $S(u_1,u_2,l,t)$ in each common bin $(l,t)$ is:
\begin{equation}\nonumber
\dfrac{P[A_1(u_1,l,t)=a_1\wedge A_2(u_2,l,t)=a_2|\sigma_I(u_1)=u_2]}{P[A_1(u_1,l,t)=a_1]\cdot P[A_2(u_2,l,t)=a_2]}
\end{equation}
where $A_i(u_i,l,t)$ is the number of actions in the given bin $(l,t)$ of $u_i$, $P[\cdot]$ is the likelihood, and $\sigma_I(u_1)=u_2$ means $u_1$ and $u_2$ are the same user.

\textbf{DG}: The third method is proposed by \cite{Chen2017Exploiting}, where a density-based
clustering method is used to extract the stay regions of a user, and a Gaussian Mixture Model (GMM) based approach is proposed to model users' temporal behaviors. Then, the similarity between two user accounts is measured based on these features.

\textbf{GS}: The fourth method is a variant of the approach proposed by \cite{Cao2016Automatic}. Based on the idea of \cite{Cao2016Automatic}, $\{(g_1,o_1),\cdots,$ $(g_m,o_m)\}$ is used to denote the observed co-occurrences of two users, where $o_i~(1\leq i\leq m)$ denotes the corresponding frequency, and the weight of $g_i$ is defined as:
\begin{equation}\nonumber
\omega(g_i) = f_s(o_i)= \dfrac{\eta}{1+e^{-\gamma o_i}} - \dfrac{\eta}{2}
\end{equation}
where $\eta$ and $\gamma$ are set to 16 and 0.2, respectively \cite{Cao2016Automatic}. Then, we can give the similarity $S(u_1, u_2)$ as follows:
\begin{equation}\nonumber
S(u_1,u_2)=\sum\limits_{i=1}^{m}\omega(g_i)\cdot\dfrac{o_i}{|R_{u_1}|}\cdot\dfrac{o_i}{|R_{u_2}|}
\end{equation}

\textbf{EEUL}: This method is our previous work \cite{Chen2018Effective}, where each user account is only represented by a sequences of grid cells, and a $k\times k$ square region is constructed to improve the efficiency.

\textbf{HFUL}: Our proposed framework HFUL differs from EEUL with following peculiarities: 1) user account similarity is measured from the spatio-temporal perspective instead of spatial perspective; 2) efficiency, effectiveness, scalability, robustness, and application are investigated; 3) an outlier detection method is designed to improve the effectiveness; 4) a novel pruning strategy is developed to reduce the number of user account pairs to be measured.

\begin{figure*}
	\centering
	\includegraphics[width=0.99\textwidth]{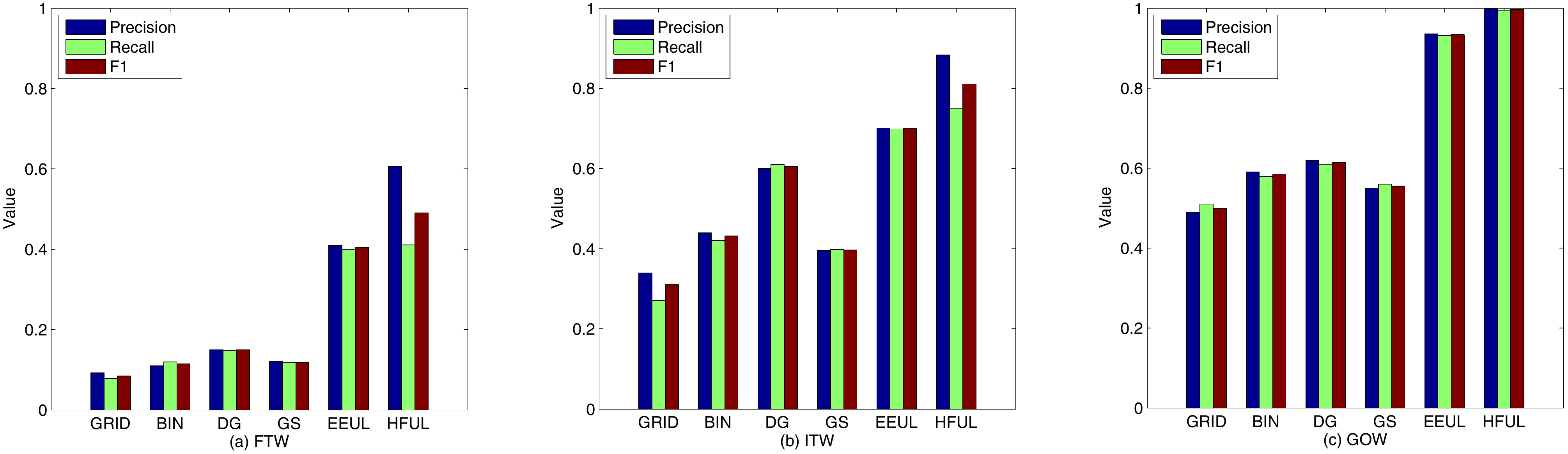}\\
	\caption{Performance of all approaches on different datasets}
	\label{7:performance of all methods}
	\vspace{-10pt}
\end{figure*}

\subsection{Evaluation Metrics}
To evaluate the effectiveness of above algorithms, we use precision, recall, and F1. Given two sets of user accounts $U_1=\{{u_1^1, u_1^2,\cdots,u_1^{m_1}}\}$ and $U_2=\{u_2^1, u_2^2,\cdots,$ $u_2^{m_2}\}$, we return the user account pair $(u_1^i, u_2^j)$ with $S(u_1^i, u_2^j)\geq S_\Delta$. The precision is defined as the fraction of user account pairs contained by the returned result that are correctly linked, and the recall is defined as the fraction of the actual linked user account pairs contained by the returned result \cite{Chen2017Exploiting},
\begin{equation}\nonumber
\begin{small}
\begin{aligned}
Recall =\dfrac{\mathcal{M}}{\mathcal{N}},Precision=\dfrac{\mathcal{M}}{\mathcal{K}},F1 =\dfrac{2\times Recall\times Precision}{Recall+Precision}
\end{aligned}
\end{small}
\end{equation}
where $\mathcal{N}$ is the number of actually linked user account pairs in the ground truth, $\mathcal{K}$ is the number of returned user account pairs, and $\mathcal{M}$ is the number of actually linked user account pairs in the returned result. To evaluate the efficiency of the proposed algorithms, we compare the time cost of them. Note that we report the best performance of baseline methods GRID, BIN, DG, GS, and EEUL on all datasets.

\subsection{Effectiveness Evaluation}
The performances of different methods are reported in Fig. \ref{7:performance of all methods}, where the precision, recall, and F1 are presented. As expected, all methods perform better than GRID, since only the common grid cells with high density are considered while measuring the similarity between a user account pair in GIRD. Both methods BIN and DG do not perform well as they did in \cite{Riederer2016Lingking}\cite{Chen2017Exploiting}. This is because we have proposed a novel evaluation metric, where all user account pairs with similarity larger than $S_\Delta$ are returned. Such metric makes our approach become more general and applicable to many applications, especially when two datasets have different numbers of user accounts and there exist many-to-many mappings. Compared with other baseline methods, our previous work EEUL performs much better, due to the following reasons. On the one hand, the kernel density estimation based similarity measurement is able to tackle the challenges data sparsity and imbalance introduced in Section \ref{1:introduction}, since we use a set of grid cells with corresponding probability to denote a user. On the other hand, we calculate the grid cell weight based on Renyi entropy, where the important and discriminative grid cells visited by few visitors are highlighted. In contrast, the popular grid cells with large entropy are assigned with small weights, due to the low discrimination of them. Without surprise, the framework HFUL performs better then EEUL, this is because: 1) we have filtered the noisy records before similarity calculation based on the outlier detection method; 2) the user account similarity is measured from the spatio-temporal perspective instead of calculating the similarity only with spatial information; 3) we reduce the computational complexity of HFUL, where the number of candidates is reduced from $m^2$ to $mk$ based on Algorithm 1.

Additionally, we find that all methods have better performance on the dataset GOW. This is because, user accounts on GOW contain more check-in records, which means the extracted features are more likely to reflect the real behaviors of a user, and we can find the actually linked pairs more precisely.

\vspace{-10pt}
\begin{table}[h]
	\centering
	\caption{Average running time (s) of different methods}
	\label{sec7:efficiency}
	\begin{tabular}{p{1cm}|p{0.8cm}|p{0.7cm}|p{0.7cm}|p{0.7cm}|p{0.7cm}|p{0.8cm}} \hline
		& GRID  & BIN  & DG & GS & EEUL & HFUL \\ \hline
		FTW	 & 24.77 & 4.01  & 3.13 & 2.36 & 0.251 & \textbf{0.201} \\ \hline
		ITW  & 120.1 & 46.63 & 4.58 & 3.57 & 0.232  &\textbf{0.224} \\ \hline
		GOW & 150.29 & 62.33 & 23.28 & 18.32 & 1.512  & \textbf{1.438} \\ \hline
	\end{tabular}
	\vspace{-20pt}
\end{table}

\begin{table*}
	\caption{Statistics of synthetic datasets}
	\centering   
	\begin{tabular}[H]{p{2.26cm}|c|c|c|c|c|c}   \hline
		& FTW & FTW2 & FTW3 & FTW4 & FTW5 & FTW6 \\ \hline
		(User, Check-in)  & (862,187795) & (1724,190780) & (2586,273346) & (3448,361741) & (4310,461745) & (5172,553501) \\ \hline
		& ITW & ITW2 & ITW3 & ITW4 & ITW5 & ITW6 \\ \hline
		(User, Check-in)  & (1717,785300) & (3434,879785) & (5151,1141451) & (6868,1582324) & (8585,1979587) & (10302,2329730) \\ \hline
		& GOW & GOW2 & GOW3 & GOW4 & GOW5 & GOW6 \\ \hline
		(User, Check-in)  & (1855,2097885) & (3710,4133293) & (5565,6213546) & (7420,8085150) & (9275,10284052) & (11130,11180847) \\ \hline
	\end{tabular}
	\label{sec10:synthesized dataset}
\end{table*}
\begin{table*}
	\caption{Time cost of HFUL on synthetic datasets}
	\centering   
	\begin{tabular}[H]{l|c|c|c|c|c|c}   \hline
		       &    FTW  &    FTW2  &    FTW3  &   FTW4   &   FTW5   & FTW6 \\ \hline
		Time-Preprocessing  & 00:02:27 & 00:02:36 & 00:02:47 & 00:02:49 & 00:02:59 & 00:03:03 \\ \hline
		Time-Calculation & 00:00:27 & 00:00:29 & 00:00:39 & 00:00:51 & 00:01:04 & 00:01:16 \\ \hline
		Time-Total       & 00:02:54 & 00:03:05 & 00:03:26 & 00:03:40 & 00:04:03 & 00:04:19 \\ \hline
		Time-Average     & 0.201s   & 0.107s   & 0.08s   & 0.064s   & 0.056s   & 0.05s \\ \hline
		       &    ITW  &   ITW2   &    ITW3  &   ITW4   &    ITW5  & ITW6 \\ \hline
		Time-Preprocessing  & 00:02:43 & 00:02:45 & 00:02:47 & 00:02:55 & 00:02:59 & 00:03:07 \\ \hline
		Time-Calculation & 00:03:42 & 00:03:57 & 00:04:48 & 00:05:24 & 00:06:00 & 00:06:33 \\ \hline
		Time-Total       & 00:06:25 & 00:06:42 & 00:07:35 & 00:08:19 & 00:08:59 & 00:09:40 \\ \hline
		Time-Average     & 0.224s   & 0.117s   & 0.089s   & 0.0732s   & 0.063s   & 0.056s \\ \hline
		       &    GOW  &    GOW2  &   GOW3   &   GOW4   &    GOW5  &   GOW6 \\ \hline
		Time-Preprocessing  & 00:26:05 & 00:26:25 & 00:26:50 & 00:27:40 & 00:28:30 & 00:28:53 \\ \hline
		Time-Calculation & 00:19:22 & 00:30:47 & 00:46:43 & 01:03:26 & 01:13:50 & 01:29:41 \\ \hline
		Time-Total       & 00:44:27 & 00:57:12 & 01:13:33 & 01:31:06 & 01:42:20 & 01:58:34 \\ \hline
		Time-Average     & 1.438s   & 0.925s   & 0.793s   & 0.736s   & 0.662s   & 0.642s \\ \hline
	\end{tabular}
	\label{sec10: total time}
	\vspace{-5pt}
\end{table*}

\subsection{Efficiency Evaluation}
The running time is another important factor needed to be considered. As seen from Table \ref{sec7:efficiency}, we report the average running time of all methods on different datasets. Obviously, GRID is the most time consuming method, as it needs to take all historical records into account while computing the density of a grid cell. Calculating the density for all grid cells before returning the top-$p\%$ ones leads to the large time cost of the method. The second method BIN is also time consuming, since we need to measure user similarity in each bin $(l,t)$. For DG, we need to spend much time to extract stay regions and time clusters, especially the weight calculation of these features. Although GS uses a set of grid cells to represent a user, the number of user accounts to be measured is $m^2$, thus the time cost of this method is larger than that of EEUL and HFUL. EEUL performs better than other baseline approaches, since a $k\times k$ square region has been constructed to prune search space. Our framework HFUL needs the least time to find an actually linked user account pair, and the main reason behind this is the reduction of computational complexity, as discussed in Section \ref{sec6:complexity analysis}. This result also demonstrates that the time cost to retrieve candidates in Algorithm \ref{sec6:candidate retrieve} has low effect on the efficiency of our propose method. That is to say, the running time saved from reducing the number of candidates to be considered is larger than that spent finding neighbors for each user account.

\subsection{Scalability Evaluation}\label{sec10:scalability evaluation}
To investigate the scalability of the proposed framework, we design a Gaussian distribution based data generator to synthesize several new datasets based on the idea of \cite{Theodoridis1999On}. For the dataset FTW, we first randomly select a user account pair $(u_1^i,u_2^i)$ from the ground truth. Then, we randomly choose several records (in the range [2,10]) from $R_{u_1^i}$ as centers to generate new records. Note that, the reason behind this selection is that \cite{Yuan2013WhoWW} has claimed that an individual's mobility usually centers at some personal geographical regions. The Gaussian probability density function of generated records around each center is $\dfrac{1}{2\pi\delta_1\delta_2}\exp\big(-\dfrac{1}{2}(\dfrac{(x-mean_x)^2}{\delta_1^2}+\dfrac{(y-mean_y)^2}{\delta_1^2})\big)$, where $mean_x$ and $mean_y$ denote the latitude and longitude of a selected record respectively, both $\delta_1$ and $\delta_2$ are set to 0.01, and based on this value the six sigma of Gaussian distribution contains 5$\times$5 grid cells centered on $(mean_x,mean_y)$. To generate the time-stamps, we use the Gaussian function $\dfrac{1}{\sqrt{2\pi}\delta}\exp\big(-\dfrac{(x-mean_t)^2}{2\delta^2}\big)$, where $mean_t$ is the time-stamp of the selected record, $\delta$ is set to 30, and the six sigma of the Gaussian distribution contains 5 time periods centered on $meant_t$. Additionally, the number of generated records of a new user account equals to $|R_{u_1^i}|$. On platform Twitter, we generate a new user count based on $u_2^i$ with similar method. Repeating the action 1724 times, we obtain the new cross-platform dataset FTW2, which contains 1724 user accounts with 190780 records. The statistics of all synthesized datasets are presented in Table \ref{sec10:synthesized dataset}.

\begin{figure*}
	\centering
	\includegraphics[width=1.0\textwidth]{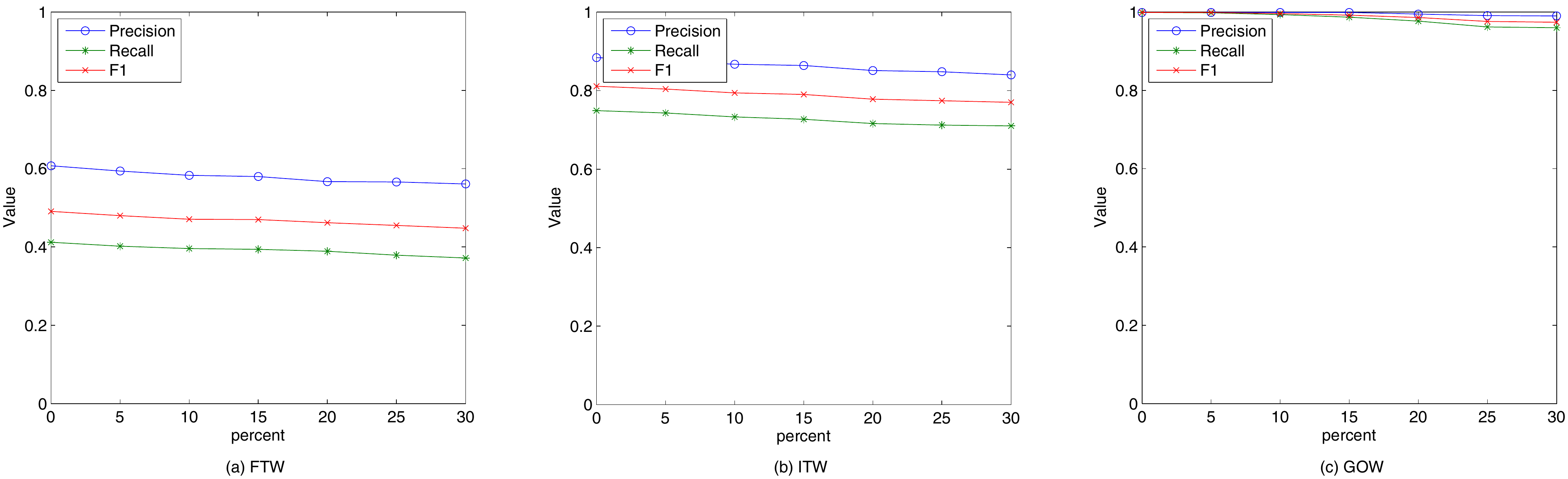}\\
	\caption{Robustness of the proposed framework HFUL}
	\label{sec7:robust}
\end{figure*}

In Table. \ref{sec10: total time}, the total time cost of HFUL on each dataset contains the time spent in preprocessing (index construction, outlier detection, feature weight calculation, and candidate retrieval) and calculating user account similarity. Without surprise, we spend more time with the increase of the size of input dataset, since more user accounts should be considered. But an interesting observation is that the time increases slowly. This is because: 1) most of preprocessing time is spent constructing $d\times d$ grid maps and $M$ time periods, and this action is same for all datasets; 2) the filtration of user accounts before computing similarity leads to the slow increase of running time. These factors lead to the decrease of the average time cost of HFUL. Additionally, we observe that the efficiency of HFUL is mainly determined by: 1) preprocessing, if the given dataset is small such as FTW-FTW6; 2) preprocessing and calculation, if the given dataset is medium such as ITW-ITW6 and GOW-GOW2; 3) calculation, if the given data is very large, such as GOW3-GOW6. 

\subsection{Robustness Evaluation}
To study the robustness of HFUL, we randomly select $5\%$, $10\%$, $15\%$, $20\%$, $25\%$, and $30\%$ records for each user account on FTW, ITW, and GOW to add noise. Specifically, for each selected record, we use the same Gaussian function in Section \ref{sec10:scalability evaluation} to generate a new record containing latitude, longitude, and time-stamp. By replacing each selected record with a new one, we can obtain some datasets with noise.

The experimental results based on these datasets are presented in Fig. \ref{sec7:robust}, observed from which the precision, recall, and F1 of HFUL will decrease while varying percentage from $5\%$ to $30\%$. Fortunately, these metrics have no significant change. This is because most of abnormal records are filtered before computing user account similarity based on the outlier detection method proposed in Section \ref{sec6:outlier detection}. In a word, the results in Fig. \ref{sec7:robust} demonstrate the high robustness of our proposed framework HFUL.

\begin{figure*}
	\centering
	\includegraphics[width=1.0\textwidth]{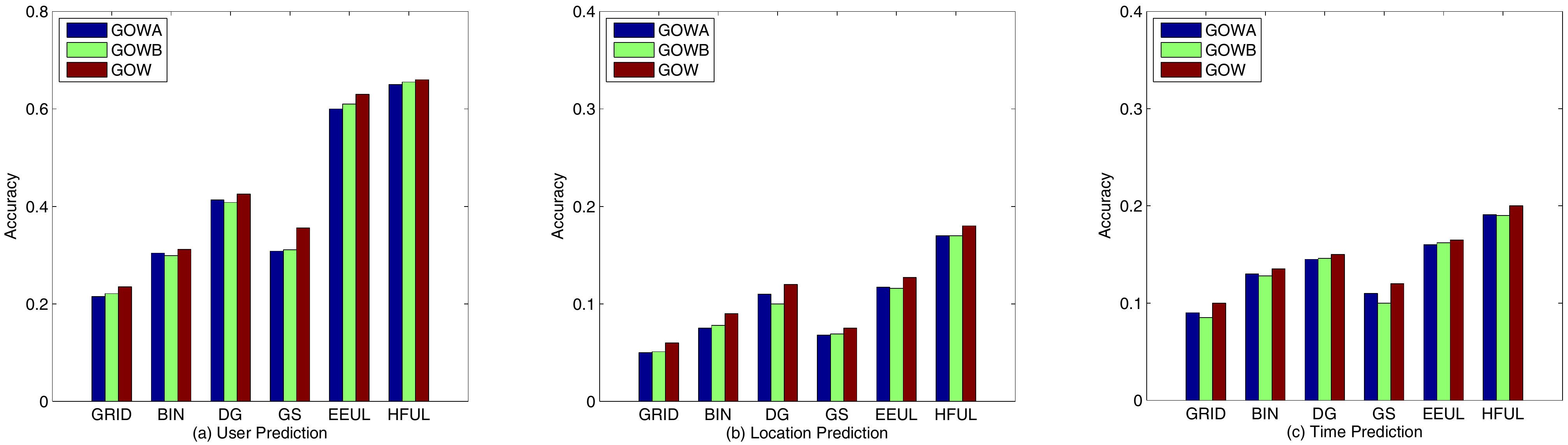}\\
	\caption{Prediction performance of all approaches on GOW}
	\label{sec7:prediction performance}
	\vspace{-5pt}
\end{figure*}

\subsection{Application Evaluation}
Following the account linkage, each user can obtain more data, based on which we study user, location, and time prediction, by choosing a 80-20 split on these data for features extraction and prediction. The results are presented in Fig. \ref{sec7:prediction performance}, where we only present the performance of HFUL on dataset GOW since the density of which is the highest, i.e., users' features extracted from which are most likely to reflect the real behaviors of them in real life. Observed from Fig. \ref{sec7:prediction performance}: 1) all algorithms have better performance on GOW, since GOWA and GOWB are only part of GOW and the features extracted from them are less likely to reflect the real behaviors of a user; 2) our proposed framework HFUL performs better than all baseline methods, this is because the linking precision and recall of HFUL are much higher than that of others.

\subsection{Impact of Different Factors}
To explore the benefits brought by outlier detection, feature weight calculation, and pruning strategy, we design following compared method: HFUL-S1, HFUL-S2, and HFUL-S3, and the properties of them are presented in Table \ref{sec7:baseline methods}.
\vspace{-5pt}

\begin{table}[h]
	\centering
	\caption{Properties of compared methods}\label{sec7:baseline methods}
	\begin{tabular}{c|c|c|c}
		\hline
		& Outlier detection & Feature weight & Pruning   \\ \hline
		HFUL-S1	& $\times$  & $\times$  & $\times$ \\ \hline
		HFUL-S2	& $\surd$   & $\times$  & $\times$  \\ \hline
		HFUL-S3	& $\surd$   & $\surd$   & $\times$   \\ \hline
		HFUL	& $\surd$   & $\surd$   & $\surd$ \\ \hline
	\end{tabular}
	\vspace{-25pt}
\end{table}

\begin{table}[h]
	\centering
	\caption{Performance of compared methods on GOW}
	\label{7:impact on performance}
	\begin{tabular}{c|c|c|c|c}
		\hline
		& HFUL-S1 & HFUL-S2 & HFUL-S3 & HFUL \\ \hline
		Precision & 0.772 & 0.786 & 0.802 & 0.999  \\ \hline
		Recall	  & 0.763 & 0.791 & 0.81 & 0.995  \\ \hline
		F1        & 0.767 & 0.788 & 0.806 & 0.997  \\ \hline
		Time cost & 36.23s & 37.68s & 40.82s & 1.438s \\ \hline
	\end{tabular}
	\vspace{-10pt}
\end{table}

\begin{figure*}
	\centering
	\includegraphics[width=1.0\textwidth]{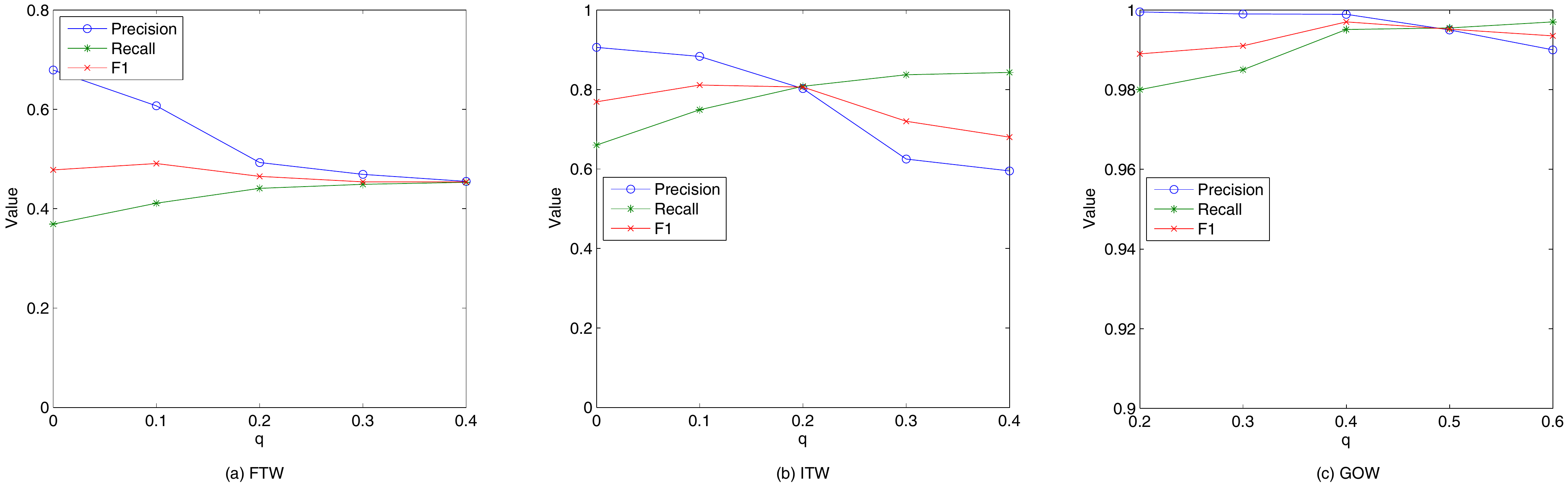}\\
	\caption{Performance of HFUL w.r.t. varied $q$}
	\label{sec7:varying q}
\end{figure*}

\begin{figure*}
	\centering
	\includegraphics[width=1.0\textwidth]{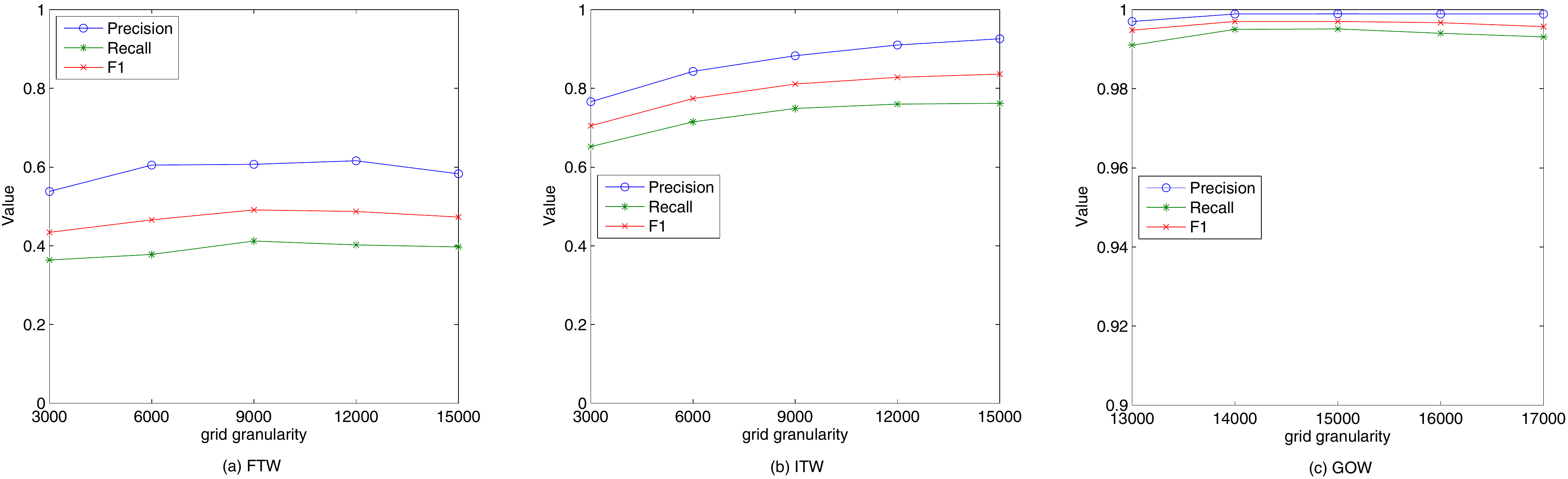}\\
	\caption{Performance of HFUL w.r.t. varied grid granularity}
	\label{sec7:varying grid granularity}
	\vspace{-10pt}
\end{figure*}

The impacts of different factors on effectiveness and efficiency are presented in Table  \ref{7:impact on performance}, where we observe that HFUL outperforms the three baselines, indicating that HFUL benefits from synchronously considering three factors in a joint way.
Compared with HFUL-S1, HFUL-S2 has lower efficiency and higher effectiveness, since it needs to spend some time to prune abnormal grids and time intervals before measuring the similarity between two user accounts. The method HFUL-S3, which benefits from feature weight calculation, is more effective than HFUL-S1 and HFUL-S2. However, the time cost of HFUL-S3 is larger than that of others, as it needs to calculate the weight for each grid cell and time interval. The highest effectiveness and efficiency of HFUL demonstrate that filtering user pairs that cannot be results with the pruning strategy not only reduces the time cost but also makes the returned results more accurate.

\subsection{Impact of Parameters}
To obtain the best performance of HFUL, tuning parameters, such as $q$, bandwidth $h$, grid granularity, number of time periods and neighbors, and the similarity threshold $S_\Delta$, is of critical importance. We therefore study the impact of different parameters in this section.

\textbf{Varying $q$.} As discussed in Section \ref{sec8:feature weight}, the elegance of using the Renyi entropy lies inside the parameter $q$. According to Eq. (\ref{sec8:grid cell weight}) and (\ref{sec8:time period weight}), we can obtain a larger and larger grid cell weight $\omega(g)$ and time period weight $\omega(T)$ with the increase of $q$. Then, it leads to the increase of user account similarity when other parameters are fixed. Just like the results of varying bandwidth $h$, the precision and recall have opposite change in Fig. \ref{sec7:varying q}, and the reasons behind the phenomenon are similar. To balance precision and recall, we set $q=0.1$ for FTW and ITW, and $q=0.4$ for GOW.
\vspace{-5pt}

\begin{table}[h]
	\caption{Average time cost (s) of HFUL w.r.t. varied grid granularity}
	\centering   
	\begin{tabular}[H]{c|c|c|c|c|c}
		\hline
		Dataset & \multicolumn{5}{c}{Grid granularity} \\\hline
		\multirow{2}{*}{FTW}  & 3000 & 6000 & 9000 & 12000 & 15000\\
		\cline{2-6}
		&  0.062 & 0.117 & 0.201 & 0.339 & 0.526 \\ \hline
		\multirow{2}{*}{ITW}  & 3000 & 6000 & 9000 & 12000 & 15000 \\\cline{2-6}
		& 0.073 & 0.145 & 0.224 & 0.319 & 0.475 \\ \hline
		\multirow{2}{*}{GOW} & 13000 & 14000 & 15000 & 16000 & 17000 \\\cline{2-6}
		& 1.396  & 1.436 & 1.438 & 1.485 & 1.489 \\ \hline
	\end{tabular}
	\vspace{-5pt}
	\label{sec7:time cost grid}
\end{table}

\begin{figure*}
	\centering
	\includegraphics[width=0.99\textwidth]{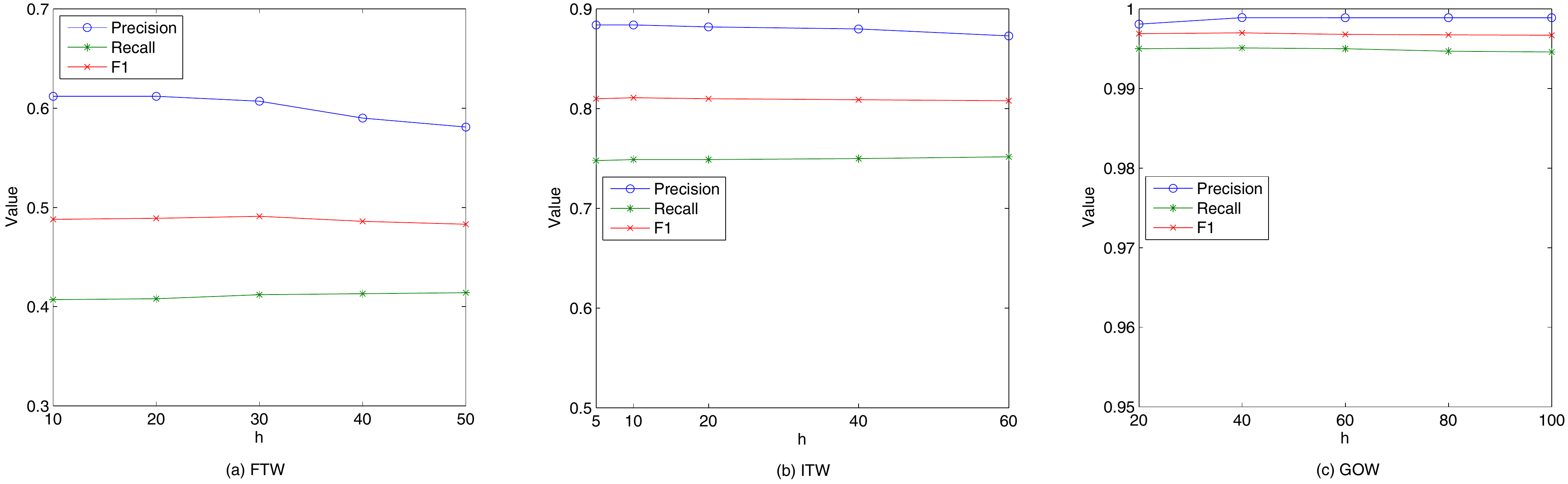}\\
	\caption{Performance of HFUL w.r.t. varied $h$}
	\label{sec7:varying h}
\end{figure*}
\begin{figure*}
	\centering
	\includegraphics[width=0.99\textwidth]{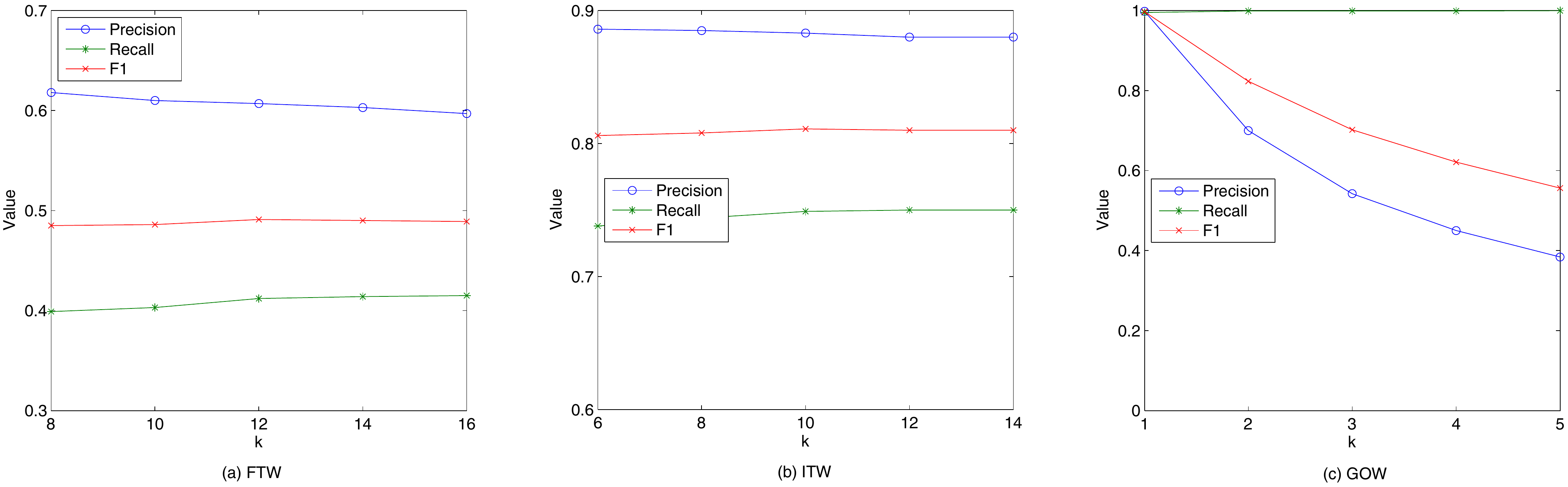}\\
	\caption{Performance of HFUL w.r.t. varied number of neighbors}
	\label{sec7:varying k value}
	\vspace{-5pt}
\end{figure*}

\textbf{Varying grid granularity.} The grid granularity is another important parameter of HFUL, where the selection of such parameter has two extremes: 1) extreme coarse granularity, the whole space is regarded as one grid cell that contains all check-in records; 2) extreme fine granularity, where each record is a grid cell and the method degrades into the naive kernel density estimation. Obviously, a too large or too small grid granularity is not appropriate for balancing effectiveness and efficiency, as presented in Fig. \ref{sec7:varying grid granularity} and Table \ref{sec7:time cost grid}. The time cost of HFUL is very sensitive to the grid granularity, since the increase of which means the records of a user may fall into more grid cells and the cardinality of grid representation of the user becomes larger. As a result, we have observed the increase of the running time of HFUL while varying the grid granularity from small to large. Taking various factors into account, we divide the space into $9000\times 9000$ for FTW and ITW, and $15000\times 15000$ for GOW. 

\textbf{Varying bandwidth $h$.} From the results in Fig. \ref{sec7:varying h}(a) and (b), we observe that the effectiveness of HFUL is sensitive to the value of bandwidth $h$, producing a larger and larger user account similarity $S(u_1,u_2)$ with the increase of $h$. Thus, it leads to the increase of recall as many user account pairs are returned, yet it also leads to the decrease of precision as many user account pairs contained by the returned result are not actually linked. As a consequence, we set $h=30m$, $h=10m$, and $h=60m$ for datasets FTW, ITW, and GOW respectively, with the goal of balancing precision and recall.

\textbf{Varying $k$}. The number of neighbors to be considered in candidate retrieval also affects the performance of HFUL. As shown in Fig. \ref{sec7:varying k value}, a too large or too small number is not appropriate for balancing precision and recall. This is because: considering too many neighbors leads to the decrease of precision, as many returned combinations are not actually linked; only considering a small number of neighbors leads to the filtration of many matched combinations. To achieve the best performance of HFUL, we set $k$ to 12, 10, and 1 for FTW, ITW, and GOW respectively. Furthermore, HFUL needs more running time with the increase of the number of neighbors since more candidates are considered, and the results are presented in Table \ref{sec8: time cost neighbor}.
\vspace{-10pt}
\begin{table}[h]
	\caption{Average time cost (s) of HFUL w.r.t. varied number of neighbors}
	\centering   
	\begin{tabular}[H]{c|c|c|c|c|c}
		\hline
		Dataset & \multicolumn{5}{c}{Number of neighbors} \\\hline
		\multirow{2}{*}{FTW}  & 8 & 10 & 12 & 14 & 16\\
		\cline{2-6}
		&  0.186 & 0.191 & 0.201 & 0.211 & 0.214 \\ \hline
		\multirow{2}{*}{ITW}  & 6 & 8 & 10 & 12 & 14 \\\cline{2-6}
		& 0.169 & 0.213 & 0.224 & 0.261 & 0.296 \\ \hline
		\multirow{2}{*}{GOW} & 1 & 2 & 3 & 4 & 5 \\\cline{2-6}
		& 1.438  & 1.815 & 2.394 & 2.887 & 3.256 \\ \hline
	\end{tabular}
	\vspace{-10pt}
	\label{sec8: time cost neighbor}
\end{table}

\begin{figure*}
	\centering
	\includegraphics[width=1.0\textwidth]{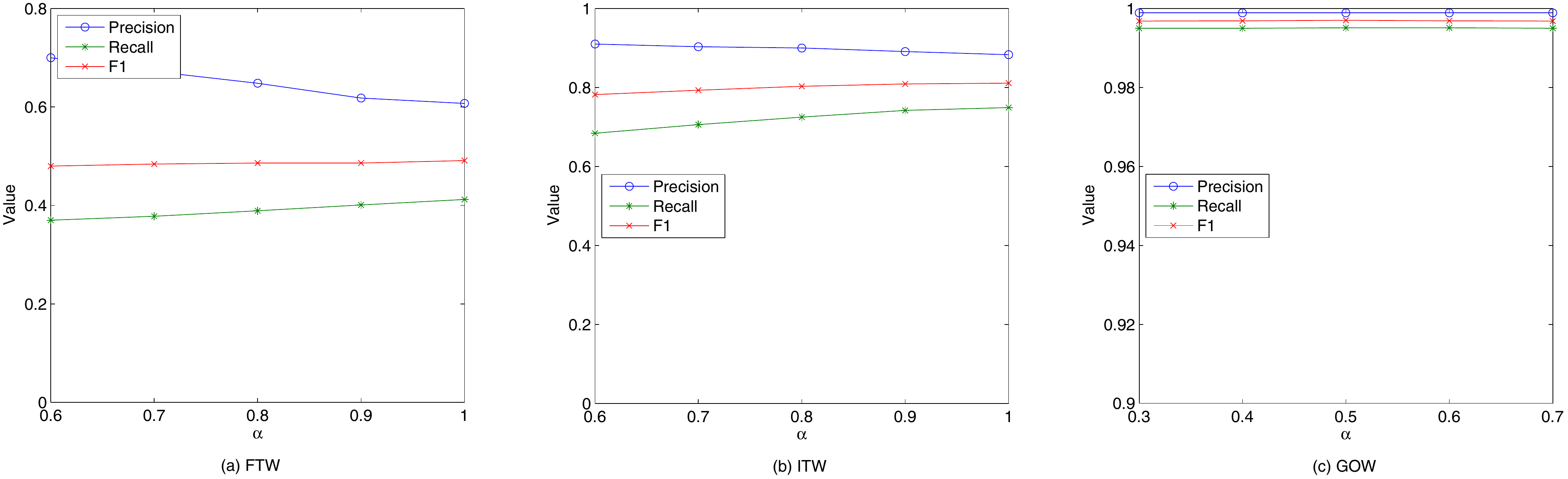}\\
	\caption{Performance of HFUL w.r.t. varied $\alpha$}
	\label{sec7:varying alpha}
\end{figure*}

\begin{figure*}
	\centering
	\includegraphics[width=1.0\textwidth]{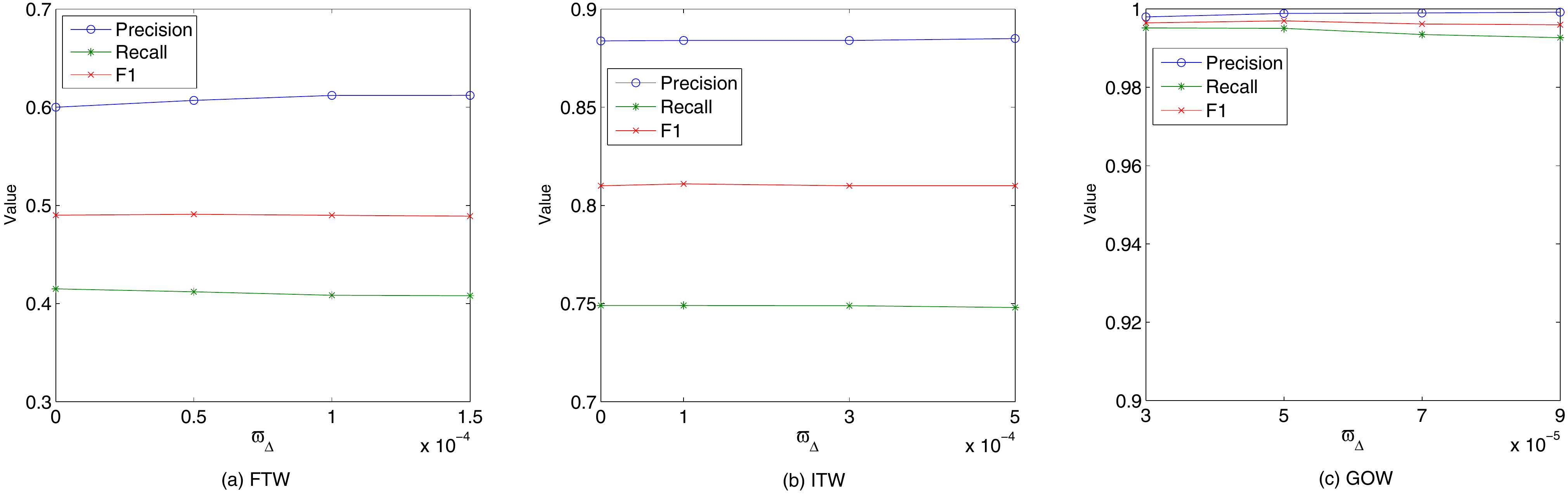}\\
	\caption{Performance of HFUL w.r.t. varied $\varpi_\Delta$}
	\label{7:varying varpi}
	\vspace{-10pt}
\end{figure*}

\textbf{Varying $\alpha$}. To study which one of spatial and temporal information is more important in linking user accounts, we vary the parameter $\alpha$ in different datasets. Observed from Fig. \ref{sec7:varying alpha}, HFUL achieves the best performance when we set $\alpha$ to 1 on datasets FTW and ITW, which means the spatial information are far more important than the temporal information in reflecting users' real behaviors on these datasets. Furthermore, HFUL has the best performance when $\alpha$ is set to 0.5 on GOW, since the quality of which is high in both spatial and temporal domain, where the user accounts belonging to the same individual have many common grids and time intervals.

\textbf{Varying $\varpi_\Delta$.} During the detection of outliers, the probability threshold $\varpi_\Delta$ is set to prune a grid cell $g$ on condition that it has no neighbor and $\varpi(g)<\varpi_\Delta$. Observed from Fig. \ref{7:varying varpi}, the precision of HFUL shows a increasing tendency on all datasets, while the recall will decrease with the increase of  $\varpi_\Delta$. This is because: given a larger $\varpi_\Delta$, more outliers are pruned, then the similarity between two specific user accounts will decrease as less grids are taken into account. Although the returned results are less likely to contain wrong pairs with a large $\varpi_\Delta$, some actually linked pairs are pruned by mistake, since many grids that are not outliers will be deleted in this case. As a result, we set $\varpi_\Delta=0.00005$, $\varpi_\Delta=0.0001$, and $\varpi_\Delta=0.00005$ for FTW, ITW, and GOW respectively, to balance the precision and recall.

\begin{figure*}
	\centering
	\includegraphics[width=1.0\textwidth]{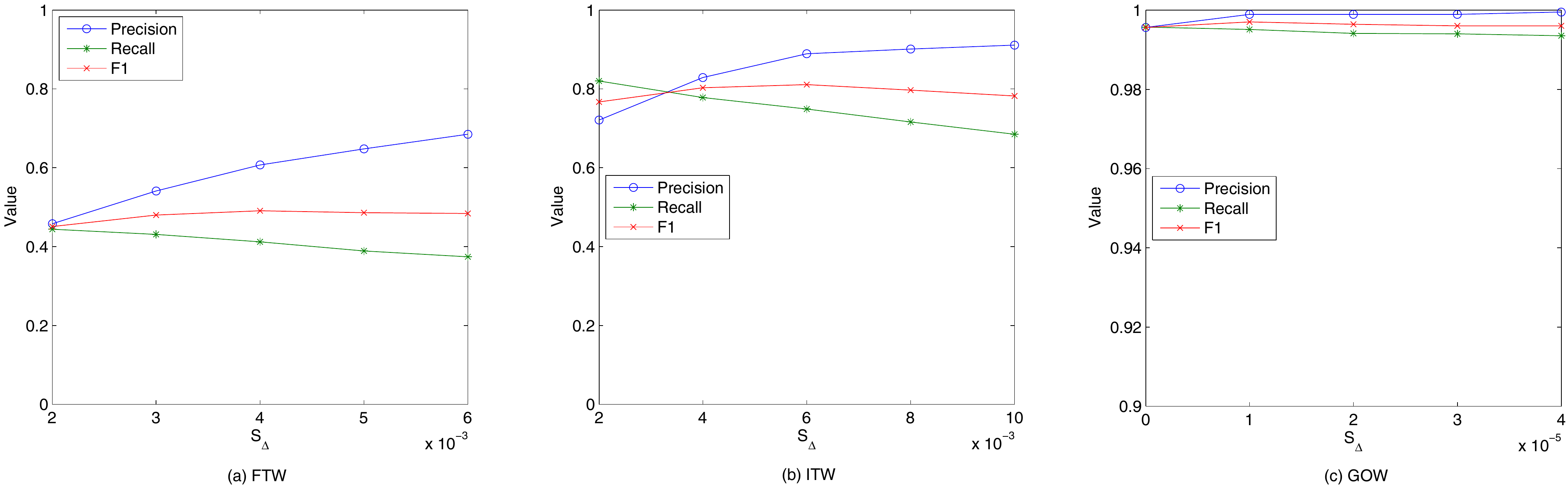}\\
	\caption{Performance of HFUL w.r.t. varied similarity threshold $S_\Delta$}
	\label{7:varying similarity threshold}
\end{figure*}
\begin{figure*}
	\centering
	\includegraphics[width=1.0\textwidth]{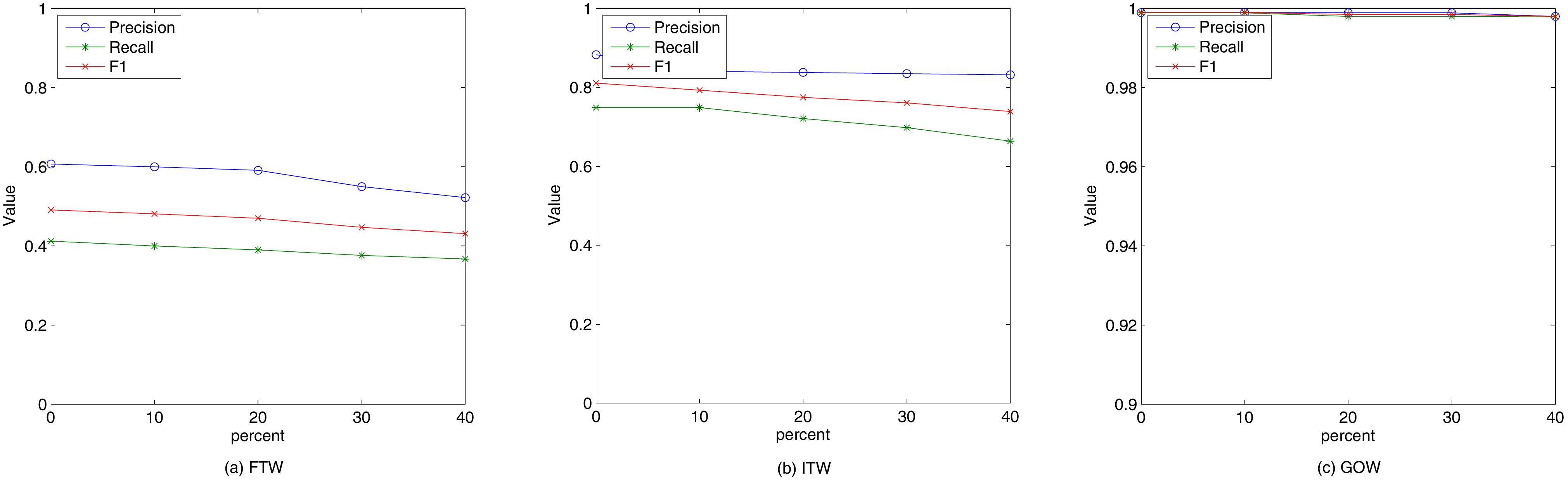}\\
	\caption{Performance of HFUL w.r.t. varied percent of biased check-ins}
	\label{7:biased percent}
\end{figure*}
\begin{figure*}
	\centering
	\includegraphics[width=1.0\textwidth]{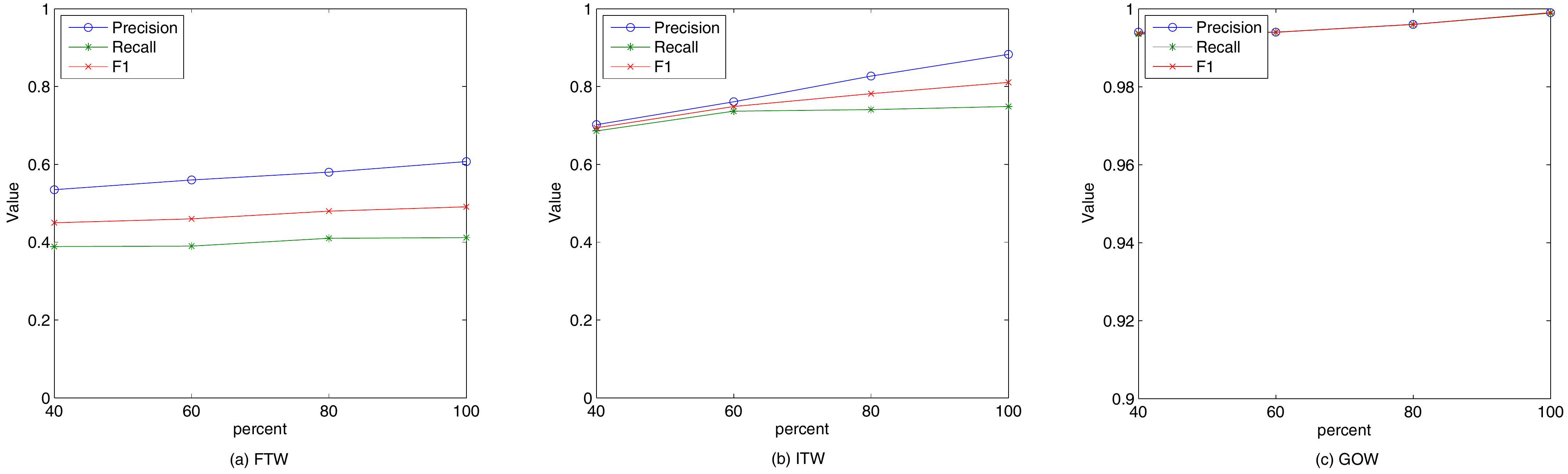}\\
	\caption{Performance of HFUL w.r.t. varied percent of original check-ins}
	\label{7:data size}
	\vspace{-5pt}
\end{figure*}

\textbf{Varying $S_\Delta$.} In real scenarios, the datasets across different platforms may have different numbers of user accounts and there may exist many-to-many mappings, thus we propose a general method where the user account pairs $\{(u_1^i,u_2^j)|u_1^i\in U_1,u_2^j\in U_2\}$ with $S(u_1^i,u_2^j)\geq S_\Delta$ are returned. Observed from Fig. \ref{7:varying similarity threshold}, the effectiveness of our method is very sensitive to the selection of $S_\Delta$. On one hand, many actual linked user account pairs are filtered with a too large $S_\Delta$. On the other hand, the returned results may contain too many unmatched user account pairs if given a small $S_\Delta$. To balance the precision and recall, and consider the characteristics of different datasets, we set $S_\Delta=0.004$, $S_\Delta=0.006$, and $S_\Delta=0.00002$ for FTW, ITW, and GOW respectively.

\textbf{Varying percent of biased check-ins}. The check-ins collected from social networks may deviate from the real locations of a user due to the instability of the GPS devices. To study the performance of HFUL in dealing with biased data, we randomly select 10\%, 20\%,30\%, and 40\% records for each user account and replace these records with that generated by the Gaussian function in Section 9.6. The results are presented in Fig. \ref{7:biased percent}, observed from which, with more records are replaced by biased check-ins, the precision, recall, and F1 of HFUL presents a downward trend. The expected decreasing trend is caused by the noise information brought by the Gaussian function. Fortunately, the performance change is not large, and this demonstrates that our proposed framework HFUL is able to handle biased check-in records.

\textbf{Varying percent of original check-ins}. To investigate the performance of HFUL in dealing with datasets with different sizes, we select 40\%, 60\%, 80\%, and 100\% original data for each user account, and the corresponding results are presented in Fig. \ref{7:data size}. Without surprise, with more check-ins are selected, the precision, recall, and F1 of HFUL presents a upward trend. This is because the similarity between two account can be measured more precisely with more abundant data.

\begin{figure}[h]
	\centering
	\includegraphics[width=0.495\textwidth]{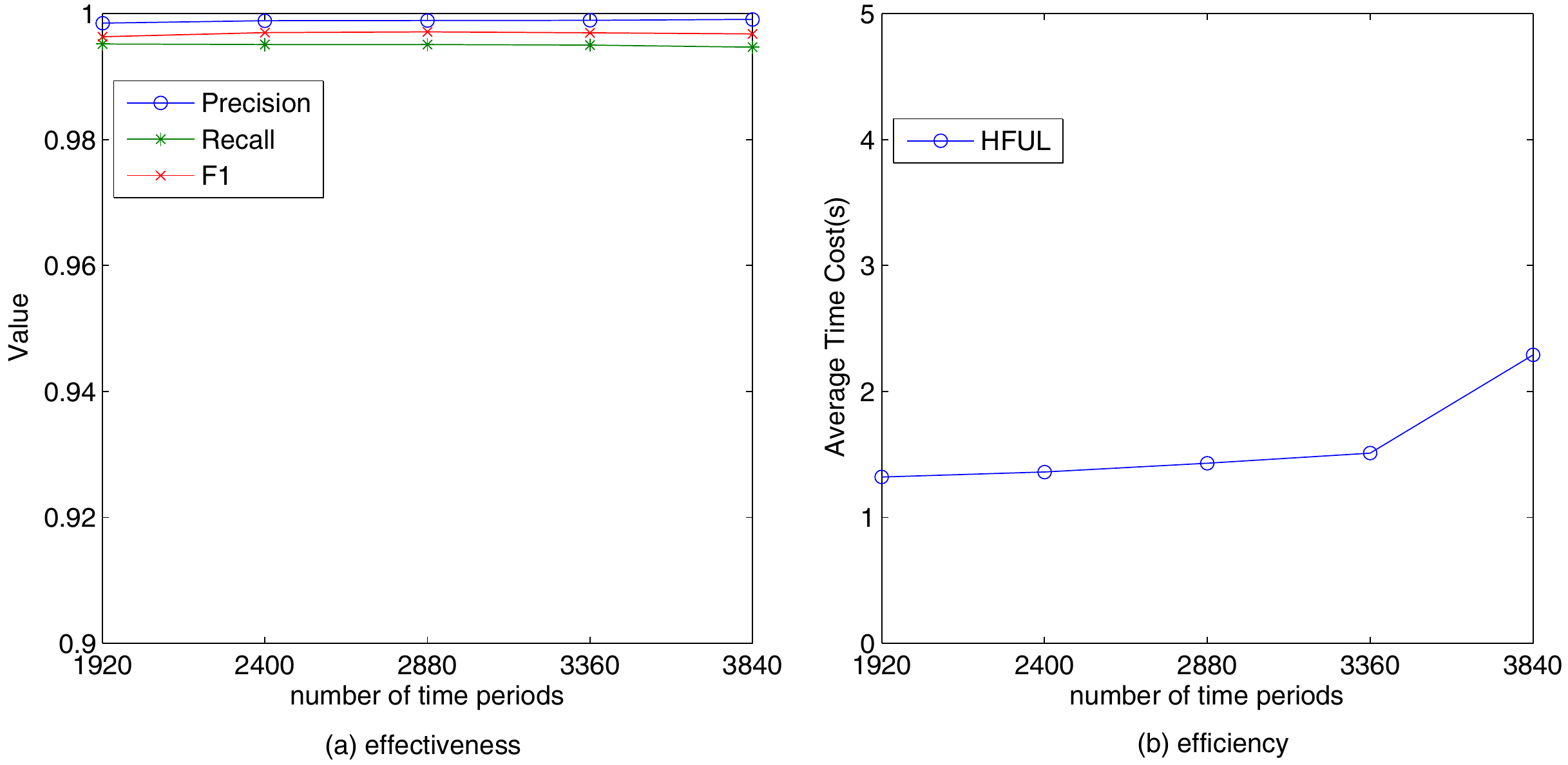}\\
	\caption{Performance of HFUL w.r.t. varied number of time periods}
	\label{sec7:varying interval}
	\vspace{-20pt}
\end{figure}

\textbf{Varying period number}. As shown in Fig. \ref{sec7:varying interval}, the temporal information is a negative factor for user account linkage on datasets FTW, and ITW, since this part of information is very sparse and many accounts belonging to same users have totally different check-in timestamps, even though their records have similar distribution in spatial domain. Thus, we only report the performance of HFUL on GOW while varying the number of time intervals from 1920 to 3840 in Fig. \ref{sec7:varying interval}. Obviously, a too small or large number is not the optimal choice. Additionally, HFUL needs to spend more time to link user accounts with the increase the number of intervals. Taking various factors into consideration, we divide temporal space into 2880 intervals.  
\vspace{-10pt}

\section{Conclusion and Future Work}\label{sec11:conclusion}
Linking user accounts across different platforms with location data has received great attention, due to the increasing availability of spatio-temporal data with check-in information, and the wide applications of the study, such as cross-platform recommendation and advertisement. To achieve user account linkage with high effectiveness, efficiency, scalability, and robustness, we have proposed several novel methods. Firstly, to tackle the data sparsity, we develop a kernel density estimation based approach to directly measure the similarity between two user accounts. Secondly, we construct the spatial and temporal indexes to improve the efficiency of HFUL and tackle the data missing problem. Thirdly, to further improve the effectiveness of HFUL, novel methods are proposed to filter outliers and calculate the weight for each grid cell and time period, where the individual ones are highlighted with large weight, yet the popular ones visited by many users are lightened due to the low discrimination of them. The experiments conducted on three real datasets demonstrate the superiority of our propose method. In the future work, we can extend the user account linkage with location data from a certain city to a global scale, by deeply exploring the cross-city, cross-country, and cross-continental check-in behaviors. Additionally, the multimodal data such as texts, photos, videos, and social graph between users can be further utilized for more precise account linkage, by developing higher performance multimodal representation learning models.

\noindent\textbf{Acknowledgments}. This work is supported by Australian Research Council Future Fellowship (Grant No. FT210100624) and Discovery Project (Grant No. DP190101985). It is partially supported by the National Natural Science Foundation of China under Grant No. 61902270 and No. 62072125,
and the Major Program of the Natural Science Foundation of Jiangsu Higher Education Institutions of
China under Grant No. 19KJA610002.

%

\bibliographystyle{IEEEtran}
\bibliography{HFUL.bbl}

\end{document}